\documentclass{article}

\usepackage{arxiv}

\usepackage[utf8]{inputenc} % allow utf-8 input
\usepackage[T1]{fontenc}    % use 8-bit T1 fonts
\usepackage{hyperref}       % hyperlinks
\usepackage{url}            % simple URL typesetting
\usepackage{booktabs}       % professional-quality tables
\usepackage{amsfonts}       % blackboard math symbols
\usepackage{nicefrac}       % compact symbols for 1/2, etc.
\usepackage{graphicx}
\usepackage{natbib}
\usepackage{amsmath, amsthm, amssymb, mathtools}
\usepackage[font=small]{caption}
\usepackage[font=small, margin=11pt, singlelinecheck=false,justification=raggedright]{subcaption}
\usepackage{graphicx} 
\usepackage{placeins}	% Preventing floats from jumping after floatbarrier
\usepackage{arydshln}	% dashed line in array

\def\cum{{\text{cum}}}
\newcommand{\multichoose}[2]{
\left.\mathchoice
  {\left(\kern-0.48em\binom{#1}{#2}\kern-0.48em\right)}
  {\!\big(\kern-0.30em\binom{\smash{#1}}{\smash{#2}}\kern-0.30em\big)\!}
  {\left(\kern-0.30em\binom{\smash{#1}}{\smash{#2}}\kern-0.30em\right)}
  {\left(\kern-0.30em\binom{\smash{#1}}{\smash{#2}}\kern-0.30em\right)}
\right.}

\newbox\tempbox
\newdimen\tabledim
\newskip\tableleftskip
\newskip\tablerightskip
\newcommand{\tbl}[2]{%
  \setbox\tempbox\hbox{#2}%
  \tabledim\hsize  
  \advance\tabledim-\wd\tempbox
  \ifdim\tabledim>0pt
  	\divide\tabledim2
  \else
    \global\tabledim0pt
  \fi
  \global\tableleftskip\tabledim  plus .5fil
  \global\tablerightskip\tabledim plus -.5fil
  \caption{#1}
  \vspace*{0.2cm}
  {\centerline{\box\tempbox}}}

\newtheorem{theorem}{Theorem}
\newtheorem{lemma}{Lemma}

\newtheorem{definition}{Definition}
\newtheorem{example}{Example}

\title{Goodness-of-fit tests for linear non-Gaussian structural
equation models}

\date{}

\author{Daniela Schkoda and Mathias Drton\\
	TUM School of Computation, Information and Technology\\
	Technical University of Munich\\
	\texttt{daniela.schkoda@tum.de, mathias.drton@tum.de} \\
}

\renewcommand{\headeright}{}

\renewcommand{\shorttitle}{Goodness-of-fit tests for linear non-Gaussian structural equation models}

\setcitestyle{authoryear,open={(},close={)}} % Round parantheses for year in \cite

\graphicspath{{Graphics/}}

\begin{document}
\maketitle

\begin{abstract}
  The field of causal discovery develops model selection methods to
  infer cause-effect relations among a set of random variables.  For
  this purpose, different modelling assumptions have been
  proposed to render cause-effect relations identifiable.  One
  prominent assumption is that the joint distribution of the observed
  variables follows a linear non-Gaussian structural equation
  model. In this paper, we develop novel goodness-of-fit tests that
  assess the validity of this assumption in the basic setting without
  latent confounders as well as in extension to linear models that
  incorporate latent confounders.  Our approach involves testing
  algebraic relations among second and higher moments that hold as a
  consequence of the linearity of the structural equations.
  Specifically, we show that the linearity implies rank constraints on
  matrices and tensors derived from moments.  For a practical
  implementation of our tests, we consider a multiplier bootstrap
  method that uses incomplete U-statistics to estimate
  subdeterminants, as well as asymptotic approximations to the null
  distribution of singular values.  The methods are illustrated, in
  particular, for the T\"ubingen collection of benchmark data sets on
  cause-effect pairs.
\end{abstract}


\section{Introduction}
The objective of causal discovery is to discern causal
relations between the components of a random vector
$X=(X_1, \ldots, X_p)$.  Often, only data from an observational
study are available, and modelling assumptions are needed if one wishes
to move beyond inference of mere equivalence classes of causal
structures \citep[\S1.8.5]{Handbook}.  To this end, several different
assumptions have been proposed; compare
\cite{heinze:2018} or
\cite{handbook:spirtes}.  One frequently adopted assumption is the linear 
non-Gaussian structural equation model introduced by
\cite{ICALiNGAM}. It posits that each component $X_i$ is a linear
function of all its causes plus an error term $\varepsilon_i$, in formulas
\begin{align}\label{eq:introduction_LiNAGAM}
X_i = \sum_{j=1, j \neq i}^p \lambda_{ji} X_j + \varepsilon_i, \quad (i=1,\ldots,p),
\end{align}
where the coefficients $\lambda_{ji}$ are real-valued parameters, 
the errors $\varepsilon_1,\ldots,\varepsilon_p$ are independent, and
specific individual  models constrain subsets of the
coefficients $(\lambda_{ij})_{i\not=j}$ to be zero.  When all but at
most one of the errors $\varepsilon_1,\ldots,\varepsilon_p$ are
non-Gaussian results from independent component analysis yield
identifiability results
that are the basis for
numerous causal discovery algorithms 
\citep{DirectLiNGAM,Lacerda_2008,Mathias_2020} as well as for
statistical inference about causal effects \citep{Strieder_2021}.

Linear non-Gaussian models also facilitate causal discovery in
scenarios involving latent confounding \citep{Hoyer_2008}.
We will thus also consider models
that include a given number  $l$ of latent variables.  Denoting
the latent variables by $L_1,\ldots,L_l$, the models then posit that
\begin{align}\label{eq:introduction_latent_conf}
X_i = \sum_{j=1, j \neq i}^p \lambda_{ji} X_j + \sum_{j=1, j \neq i}^l  \gamma_{ji} L_j  + \varepsilon_i, \quad (i=1,\ldots,p).
\end{align}
Chapter 5 of \cite{book_shimizu} reviews progress in model selection
for this class of models.    For
further recent work we refer to
\cite{salehkaleybar:2020} and \cite{Mathias_Sam_JMLR}.

While linear structural equation models play an important role in
exploring causal structures, the conclusions they lead to may be
unreliable in the face of (significant) model misspecification.  This motivates the
work presented here, which develops a method to formally test if the
linear structural equation models
from~\eqref{eq:introduction_LiNAGAM}, or more generally,
from~\eqref{eq:introduction_latent_conf} are tenable for the data at
hand.

If the causal structure admits the form of a directed acyclic graph,
no latent confounding is present, and a causal order of the variables
is known, then the linearity assumptions could be assessed by
exploring the fit of linear regression models as models of the
conditional distribution of a variable given preceding variables;
cf.~\cite{Mathias_Sam_JMLR}.  \cite{schultheiss:2023} discuss
aspects of goodness-of-fit when no causal order is known and the
causal structure takes the form of a directed acyclic graph.  In
contrast, our interest is in a global goodness-of-fit test when
the causal structure may be entirely arbitrary.  For such scenarios,
the stability of an algorithm's output under bootstrapping
has been considered to assess the suitability of causal
discovery methods \citep{Raghu_2018,Biza_2020}.  While this idea is
generally applicable, it fails to penalize
systematic errors.  Moreover, it is difficult to arrive at a formal
inferential statement about goodness-of-fit of a model, which is the
goal of this work.

\cite{Matteson_2017b} study a joint dependence measure and its
application for estimation in the related model of independent
component analysis.
The authors also suggest that their joint dependence measure can be
applied for testing goodness-of-fit
by testing the joint independence
of error terms.
To implement this suggestion in our context, one needs to first
estimate the inverse of the linear map specified
via~\eqref{eq:introduction_LiNAGAM}.  The inverse map may then be used
to form residuals that serve as estimates of the error terms.
Treating the residuals as a sample drawn from the joint distribution
of the errors, one may estimate the joint dependence measure.
However, for a rigorous statistical test, one needs to account for the fact that residuals are only
noisy estimates of the actual errors $\varepsilon_i$ \citep{sen:sen:2014}.  \cite{Matteson_2017b} thus
suggest a bootstrap method to approximate the null distribution of the
estimated joint dependence measure for the errors.

In this paper, we develop a more direct approach to testing
goodness-of-fit of the linear model
from~\eqref{eq:introduction_LiNAGAM}.  To this end, we take an
algebraic perspective and assess how the model constrains moments.
While this rests on the assumption that the relevant moments indeed
exist, the algebraic relations we consider hold for
any distribution, Gaussian or non-Gaussian. Furthermore, our approach
naturally generalizes to the case with latent confounding, i.e., the
models from~\eqref{eq:introduction_latent_conf}.
The considered relations take the form of rank constraints on matrices and
tensors formed from the moments.
To test the constraints statistically, we consider an approach based
on estimating vanishing subdeterminants via incomplete U-statistics
as well as tests that are based on singular value decomposition.  Our
method has the advantage of avoiding repeated model fitting on
bootstrap samples, and our numerical studies show that our method is
significantly faster while reaching similar power when compared
to the method of \cite{Matteson_2017b} in most  of the considered simulation settings.

\section{Preliminaries}
\subsection{Linear Structural Equation Model} 
The linear structural equation model with latent confounding is the
set of all probability distributions $P^X$ on $\mathbb{R}^p$ that
arise as joint distributions of random vectors $X = (X_1, \ldots, X_p)$
that uniquely solve the linear system
\begin{equation}
\label{def:lin sem}
X = \Lambda^T X + \Gamma^T L + \varepsilon
\end{equation}
for a choice of coefficient matrices $\Lambda \in \mathbb{R}^{p \times
  p}$ and $\Gamma \in \mathbb{R}^{l \times p}$, and random vectors
$\varepsilon \in \mathbb{R}^p$ and $L \in \mathbb{R}^l$ with all
$p+l$ components $\varepsilon_i$ and $L_j$ mutually independent.
The matrix $\Lambda$ is taken to have all diagonal elements zero and,
for unique solvability, is such that $I - \Lambda$ is invertible. We allow $l$ to be zero, which corresponds to the case without
latent confounding.

\begin{figure}
    \centering
     \includegraphics[width=0.33\linewidth]{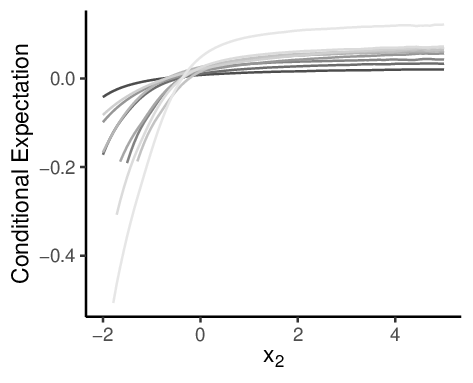}
    \caption{In a linear SEM, the conditional expectations are not necessarily linear, as exemplified here for $E(X_1\mid X_2=x_2)$ with  $\eta_1, \eta_2$ generated according to a Gamma distribution with random shapes and scales; each line reflects a different random draw. The linear coefficients are fixed to $b_{21}=b_{22}=1$.}
    \label{fig:non_linear_expectation}
\end{figure}

Subsequently, it will be convenient to rewrite the
system in~\eqref{def:lin sem} in the equivalent form
\begin{equation}
\label{Lin SEM equiv form}
X = B \eta,
\end{equation}
where $B = (I_p - \Lambda)^{-T} \begin{pmatrix}
  \Gamma^T & I_p
\end{pmatrix}$ and
$\eta=(L_1,\ldots,L_l,\varepsilon_1,\ldots,\varepsilon_p)$ comprises the
errors and latent variables.
Note that \eqref{Lin SEM equiv form} has the form of (overcomplete) independent component analysis \citep{HyvarinenKO01}.
We denote the entries of the coefficient matrix by $B =
\left(b_{ij}\right)_{i, j \in [p]}$, where $[p]=\{1,\ldots,p\}$.
Without loss of generality,  $\eta$
and consequently $X$ are assumed to have mean zero. We want to highlight that despite the linearity of the structural equations in \eqref{def:lin sem}, the conditional expectations of the different variables are not necessarily linear. For example, as shown  in Figure \ref{fig:non_linear_expectation}, if $p=2$ and $l=0$, then $E(X_1\mid X_2=x_2) = E(b_{11}\eta_1 + b_{12}\eta_2 \mid b_{21}\eta_1 + b_{22}\eta_2 = x_2)$, which is generally nonlinear if neither $b_{21}$ nor $b_{22}$ is zero. This nonlinearity prevents direct application of goodness-of-fit techniques from linear regression models.

\subsection{Tensors and Tensor Rank}
For $q,k\in\mathbb{N}$, let $(\mathbb{R}^q)^{\otimes k}$
be the $k$-fold tensor product of $\mathbb{R}^q$.  A tensor $T=
(t_{i_1\ldots i_k}) \in \left(\mathbb{R}^q\right)^{\otimes k}$ is
symmetric if $t_{i_1\ldots i_k} = t_{\pi(i_1)\ldots \pi(i_k)}$  for all
permutations $ \pi: [q] \to [q]$.  We write $
\text{Sym}_k(\mathbb{R}^q)$ for the subspace of all
symmetric tensors in  $(\mathbb{R}^q)^{\otimes k} $.  The Tucker
product of $T$ and
$k$ copies of a matrix $A=(a_{ij}) \in \mathbb{R}^{p \times q}$ is the tensor in
$(\mathbb{R}^p)^{\otimes k}$ given by
\begin{equation*}
\left(T \bullet A \bullet A \bullet \cdots \bullet A\right)_{i_1 \ldots i_k} = \sum_{j_1, \ldots, j_k = 1}^q t_{j_1 \ldots j_k} a_{i_1 j_1} \cdots a_{i_k j_k}, \quad (i_1, \ldots, i_k \in [p]).
\end{equation*}
We write $\multichoose{q}{m} = \binom{q+m-1}{m}$ for the number of
ways to choose $m$ out of $q$ elements with repetition while ignoring
the order. For $m \leq k$, the $m$th flattening of the  symmetric tensor $T \in \text{Sym}_k(\mathbb{R}^q)$ is
the $\multichoose{q}{m} \times \multichoose{q}{k-m}$ matrix 
$\text{fl}_m(T)$ whose entries are
$$\left(\text{fl}_m(T)\right)_{(i_1,\ldots, i_m),(i_{m+1},\ldots,i_k)} = t_{i_1\ldots i_k}.$$
Here, the matrix columns are indexed by $(i_1,\ldots, i_m)\in [q]^m$
with $i_1 \leq \cdots \leq i_m$, and the rows are indexed by
$(i_{m+1}, \ldots, i_k) \in [q]^{k-m}$ with
$i_{m+1} \leq \cdots \leq i_k$.  A symmetric tensor
$T \in \text{Sym}_k(\mathbb{R}^q)$ has real symmetric rank $r$ if $r$
is the smallest integer such that
\begin{align}\label{def:sym rank}
T = \bigg(\sum_{j=1}^r a_{i_1 j}\cdots a_{i_k j}\bigg)_{i_1, \ldots, i_k \in [q]}
\end{align}
for a matrix $A=(a_{ij}) \in \mathbb{R}^{q \times r}$. Moreover, $T$ has symmetric border rank $r$ if $r$ is the smallest integer such
that $T$ is in the closure of the set of tensors with symmetric rank at most $r$.
\subsection{Parametrization of the Cumulants}

We are interested in analyzing the structure of the cumulants
realizable under the linear structural equation model. Cumulants are
tensors consisting of alternating sums of moments.  Let
$Z$ be a random vector taking values in
$\mathbb{R}^q$, with joint distribution $P^Z$.  Then the
$k$th-order cumulant tensor of $P^Z$  is the tensor  $\cum^{(k)}(P^{Z}) \in
\text{Sym}_k(\mathbb{R}^q)$ 
given by
\begin{equation*}
\Big(\cum^{(k)}\big(P^{Z}\big)\Big)_{i_1 \ldots i_k} = \sum_{(I_1, \ldots, I_h)} (-1)^{h-1} (h-1)! E \left( \prod_{j \in I_1} Z_j \right) \cdots E \left( \prod_{j \in I_h} Z_j \right),
\end{equation*}
where $(I_1, \ldots, I_h)$ is an arbitrary partition of $(i_1, \ldots,
i_k)$. If $Z$  is centred, the second-order cumulant tensor is the
covariance matrix, and the third-order cumulant tensor consists of all
the third moments $E(Z_i Z_j Z_k)$; see also \citet[Chapter
2]{McCullagh_2018}.

To facilitate discussion of the cumulants of random vectors $X$ that
follow a linear structural equation model with latent confounders, we
make the following definition.

\begin{definition}
  For $l\ge 0$ latent confounders, and orders $k_1 < k_2$, the  
  cumulant
  model is the set
\begin{align*}
\mathcal{C}_l^{(k_1, \dots, k_2)} = \Bigl\{  &\left( \cum^{(k_1)}\left(P^{X}\right), \ldots, \cum^{(k_2)}\left(P^{X}\right)\right):  P^{X} \text{ is realizable}\\  & \text{ under a linear structural equation model with $l$ latent confounders} \Bigr\} .
\end{align*}
\end{definition}
Subsequently, we use the shorthand $C^{(k)}$ to denote the $k$th-order cumulant of $P^X$, and we tacitly assume that the noise vector
$\eta$ has all cumulants up to order
$k$ finite.  In reference to~\eqref{Lin SEM equiv form},
the cumulant model can be parametrized in terms of $B$ and the cumulants of $\eta$. 
\begin{lemma}\label{lem:parametrization} 
If $P^X$ satisfies a linear structural equation model with $l$ latent confounders, then
\begin{equation*}
C^{(k)} = \cum^{(k)}\left(P^{X}\right) =  \cum^{(k)}\left(P^{\eta}\right)\bullet B \bullet \cdots \bullet B,
\end{equation*}
where $B$ appears $k$ times in the Tucker product.
\end{lemma}
A proof can be found in \cite{Comon_2010}.

\section{Algebraic Structure of the Cumulants}
\subsection{Necessary Conditions across Cumulants}

In this section, we establish the theoretical underpinnings for the
test statistics in our goodness-of-fit tests. Combining the equations
from Lemma \ref{lem:parametrization} for different $k$, we show that
the rank of a
matrix formed from the cumulants drops for distributions in the linear structural equation model. 
\begin{theorem}
\label{thm:rank_cond_M}
For  $k_1 < k_2$, construct the matrix
\begin{equation*}
M^{(k_1, \ldots, k_2)} = \left(\begin{array}{c}
\quad \text{fl}_{k_1}\left(C^{(k_1)}\right)\quad \\[3pt] \hdashline
\vdots \vspace{3pt}
\\ \hdashline \\ \quad \text{fl}_{k_1}\left(C^{(k_2)}\right) \quad 
\end{array}\right)\in \mathbb{R}^{ 
\sum_{h=k_1}^{k_2} \multichoose{p}{h-k_1}\times\multichoose{p}{k_1}}
\end{equation*}
that contains the vectorized $k_1$th cumulant as first row and the
higher cumulants rearranged underneath.  If $\left(C^{(k_1)}, \ldots,
  C^{(k_2)}\right)$ lies in $\mathcal{C}_l^{(k_1, \dots, k_2)}$, then the rank
of $M^{(k_1, \ldots, k_2)}$ is at most $p+l$.
\end{theorem}

The rank condition on $M^{(k_1, \ldots, k_2)}$ is non-trivial only if
 the number of rows and columns of $M^{(k_1, \ldots, k_2)}$ exceeds the
 rank bound. The number of rows is determined by the choice of $k_1$ 
 and the number of columns by the choice of $k_2 - k_1$. To exemplify 
 this point, take $k_1=2$ and $k_2=3$, which gives
$$M^{(2,3)} =\left(
\begin{array}{cccccccc}
c^{(2)}_{11} & c^{(2)}_{12} & \cdots & c^{(2)}_{1p} & c^{(2)}_{22} & c^{(2)}_{23} & \cdots & c^{(2)}_{pp} \\[3pt] \hdashline 
 &  &  &  &  &  & & \\[\dimexpr-\normalbaselineskip+3pt]
c^{(3)}_{111} & c^{(3)}_{112} & \cdots & c^{(3)}_{11p} & c^{(3)}_{122} & c^{(3)}_{123} & \cdots & c^{(3)}_{1pp} \\
\vdots  & \vdots  & \ddots & \vdots & \vdots & \vdots  & \ddots & \vdots \\
c^{(3)}_{p11} & c^{(3)}_{p12} & \cdots & c^{(3)}_{p1p} & c^{(3)}_{p22} & c^{(3)}_{p23} & \cdots & c^{(3)}_{ppp} 
\end{array}\right) \in \mathbb{R}^{\left(1+p\right)\times \binom{p+1}{2} }.
$$
Then, with $p\ge 2$, the rank condition is non-trivial if and only if
$l=0$.  In other words, in the absence of latent confounding,
Theorem~\ref{thm:rank_cond_M} offers a way to test linearity of the
structural equation model on the basis of second and third moments.
For $l \geq 1$, we need to consider higher cumulants to determine
testable constraints. For computational and statistical reasons, we 
do not use orders higher than necessary. Simulations, detailed in the supplementary material, confirm that selecting the lowest choice possible yields superior performance. For $p\geq 3$ 
and $1 \leq l \leq p^2 + 1$, this leads to 
the choice $(k_1, k_2) = (2,4)$. For $p=2$ and $l=1$, the choice $k_1
= 2$ does not suffice since 
the number of columns is too low. Instead, $(k_1, k_2) = (3,5)$ is the lowest choice possible.  In principle, we can handle an arbitrarily high number of confounders by increasing $k_1$ and $k_2$ even more, 
but focus here on the above-mentioned cases.
\begin{proof}[of Theorem~\ref{thm:rank_cond_M}] Since the 
 vector $\eta$ has independent components, all its cumulants $\cum^{(k)}(P^\eta)$ are diagonal. Expanding the equations from Lemma \ref{lem:parametrization}, one obtains
\begin{align*}
c^{(k)}_{i_1\ldots i_k} &= \sum_{j=1}^{p+l} \Big(\cum^{(k)}(P^\eta)\Big)_{j\ldots j} (b_{i_1 j} \cdots b_{i_k j}), \quad (i_1,\ldots, i_k \in [p]).
\end{align*}
The claim now follows because all rows of $M^{(k_1, \ldots, k_2)}$ are contained in
\begin{equation*}
\text{span}\left(\left\{\begin{pmatrix}
b_{1j}b_{1j}\cdots b_{1j}\\
b_{1j}b_{1j}\cdots b_{2j}\\
\vdots   \\
b_{pj}b_{pj}\cdots b_{pj}
\end{pmatrix} : j =1, \dots, p+l \right\}\right).  
\end{equation*}
\end{proof}
\subsection{Necessary Conditions within Cumulants}
The constraints in Theorem~\ref{thm:rank_cond_M} are based on shared
structure in cumulant tensors of different order.  However,
considering each cumulant separately, one obtains a second type of
constraints, as a direct consequence of Lemma
\ref{lem:parametrization}. 
\begin{theorem} \label{thm:tensor rank} Let $k \geq 2$. If $C^{(k)}$
  lies in $\mathcal{C}^{(k)}_l$, then the tensor $C^{(k)}$ has symmetric rank at most $p+l$.
\end{theorem}

Theorem~\ref{thm:tensor rank} encompasses the case
$k=2$.  However, $C^{(2)}$ is a matrix in $\mathbb{R}^{p \times p}$ and,
thus, always has rank at most $p$.  As a result, Theorem~\ref{thm:tensor rank}
does not yield non-trivial constraints involving second moments.  Instead, we need
to consider cumulant tensors of order three and beyond, where the
symmetric tensor rank generally surpasses $p$.

For tensors of order three and higher, the
notion of rank is more delicate to work with.  In particular,
the set of tensors of rank at most $r$ is not closed, and if only a
finite sample approximation of $C^{(k)}$ is available, one can not
hope to distinguish whether $C^{(k)}$ lies in the set or its closure.
Consequently, we focus on testing the slightly relaxed condition
that $C^{(k)}$ belongs to the closure of tensors with rank at
most $p+l$.  In other words, we test if $C^{(k)}$ has symmetric border
rank at most $p+l$.  To conduct a test in practice, we will exploit
that the symmetric border rank can be related to polynomial
conditions and rank conditions, 
which are summarized in the following theorems. We start with the bivariate case without
latent confounding, so $p=2$ and $l=0$.  

 \begin{theorem}\label{thm:ineq2}
A symmetric tensor  $T=(t_{i_1i_2i_3}) \in \text{Sym}_3(\mathbb{R}^2)$ has symmetric border rank at most two if and only if
$$\text{Str}(T)=3t_{112}^2t_{122}^2-4t_{111}t_{122}^3-4t_{112}^3t_{222}+6t_{111}t_{112}t_{122}t_{222}-t_{111}^2t_{222}^2 \leq 0.$$
\end{theorem}

The theorem is derived in the supplementary material.
With one more observed variable, so $p=3$ and $l=0$, the rank condition is equivalent to a polynomial equality of \cite{aronhold}.
\begin{theorem}
\label{thm:Aronhold} A symmetric tensor $T=(t_{i_1i_2i_3}) \in
\text{Sym}_3(\mathbb{R}^3)$ has symmetric border rank at most three if
and only if $\text{Ar}(T)=0$, where $\text{Ar}(T)$ is the so-called \textit{Aronhold invariant}:
\begin{align*}
\text{Ar}(T)=\;&t_{111}t_{222}t_{333}t_{123} - (t_{222}t_{333}t_{112}t_{113} + t_{333}t_{111}t_{122}t_{223} + t_{111}t_{222}t_{133}t_{233})\\
&- t_{123}(t_{111}t_{223}t_{233} + t_{222}t_{133}t_{113} + t_{333}t_{112}t_{122}) + (t_{111}t_{122}t_{233}^2 + t_{111}t_{133}t_{223}^2
 \\ 
&+ t_{222}t_{112}t_{133}^2 + t_{222}t_{233}t_{113}^2 + t_{333}t_{223}t_{112}^2 + t_{333}t_{113}t_{122}^2)
-t_{123}^4 \\
&+ 2 t_{123}^2(t_{122}t_{133} + t_{233}t_{112} +t_{113}t_{223}) - 3 t_{123}(t_{112}t_{223}t_{133} + t_{113}t_{122}t_{233})
\\ 
& - (t_{122}^2 t_{133}^2 + t_{233}^2 t_{112}^2 + t_{113}^2 t_{223}^2) + (t_{233}t_{112}t_{113}t_{223} + t_{113}t_{223}t_{122}t_{133} \\
&+ t_{122}t_{133}t_{233}t_{112}).
\end{align*}
\end{theorem}
The theorem follows from combining Theorem 1.2 in \cite{Ottaviani} with Theorem 2.1 and Theorem 2.2 in \cite{MaurizioBanchi_2015}.

For $p=4$ and $l=0$, the border rank constraint can be characterized 
in terms of 36 quintics, which are given in Proposition 3.2 in \cite{Seigal}. However, testing for them in our
framework would be computationally expensive. So, instead, we assess a simpler to handle necessary rank constraint, 
which generalizes for arbitrary $p$. Specifically, the Young
flattening  $Y_3(T)$ of $T\in \text{Sym}_3(\mathbb{R}^p)$ drops rank
if the tensor has rank at most $p$. Let $a=\lfloor \frac{p-1}{2}
\rfloor$.  Then 
$Y_3(T)$ is the matrix of size ${p\binom{p}{a}\times p\binom{p}{a+1}}$ with entries
\begin{multline*}
\left(Y_3(T)\right)_{(i_1, \ldots, i_{a+1}), (j_1, \ldots, j_{a+2})} =\\
\begin{cases}
t_{j_1, i_1, c} \cdot \text{sgn}\left(\left(
\begin{smallmatrix}
j_2 & \ldots & j_{a+1} & j_{a+2} \\
i_2 & \ldots & i_{a+1} & c
\end{smallmatrix}
\right)\right) & \text{if } \{i_2, \ldots, i_{a+1}\} \subseteq \{j_2, \ldots, j_{a+2}\}, \\
0 & \text{otherwise}.
\end{cases}
\end{multline*}
Here,  the rows are indexed by tuples $(i_1, \ldots, i_{a+1}) \in
[p]^{a+1}$ with $ i_2 < \cdots < i_{a+1}$, the columns by tuples $(j_1, \ldots, j_{a+2}) \in [p]^{a+2}$ with $j_2 < \cdots < j_{a+2}$, and $c$ is the number satisfying $\{i_2, \ldots, i_{a+1}, c\} = \{j_2, \ldots ,j_{a+2}\}$.
\begin{theorem} \label{thm:AT} Let $p \geq 3$. If $T \in
  \text{Sym}_3(\mathbb{R}^p)$ has symmetric border rank at most $p$,
  then $Y_3(T)$ has rank at most $\binom{p-1}{a}p$.
\end{theorem}
The theorem is proven as Theorem 1.2.3 in \citet{Landsberg_2013} in a coordinate-free way. We give an explicit construction of the matrix $Y_3(T)$ in the supplementary material.
For example, for $p=3$, $Y_3(T)$ is a $9\times 9$ matrix and of  the form 
\begin{equation*}
Y_3(T) =
\begin{pmatrix}
0 & t_{113} & -t_{112} & 0 & -t_{123} & t_{122} & 0 & t_{133} & -t_{123} \\
-t_{113} & 0 & t_{111} & t_{123} & 0 & -t_{112} & -t_{133} & 0 & t_{113} \\
t_{112} & -t_{111} & 0 & -t_{122}& t_{112} & 0  & t_{123} & -t_{113}  & 0\\

0 & -t_{123} & t_{122} & 0 & t_{223} & -t_{222} & 0 & -t_{233} & t_{223} \\
t_{123} & 0 &  -t_{112} & -t_{223} & 0 & t_{122} & t_{233} & 0 & -t_{123} \\
-t_{122} &t_{112} & 0 & t_{222}& -t_{122} & 0  & -t_{223} & t_{123}  & 0\\

0 & t_{113} & -t_{123} & 0 & -t_{233} & t_{223} & 0 & t_{333} & -t_{233} \\
-t_{133} & 0 & t_{113} & t_{233} & 0 & -t_{123} & -t_{333} & 0 & t_{133} \\
t_{123} & -t_{113} & 0 & -t_{223}& t_{123} & 0  & t_{233} & -t_{133}  & 0
\end{pmatrix}.
\end{equation*} The theorem states that this skew-symmetric matrix has rank at most six. So, in particular, all the principal $8$-minors vanish. All these minors coincide up to scaling with the square of the Aronhold invariant. Hence, one of the directions of Theorem \ref{thm:Aronhold} is a consequence of Theorem \ref{thm:AT}.

In the presence of latent confounding, for $p=2$, there exist
inequalities for cumulant order four, see the Main Theorem in \cite{Comon_2012}.
For computational reasons, we do not use them but turn to
order six, where a rank constraint is fulfilled. The same result holds for $p=3$ and order four. 
\begin{theorem}
  Let $(p, k)=(2, 6)$ or $(p, k)=(3,4)$, and $1 \leq l < p$. Then  $T \in \text{Sym}_k\left({\mathbb{R}^p}\right)$ has symmetric border rank at most $p+l$ if and only its flattening $\text{fl}_{k/2}(T) \in \mathbb{R}^{k/2 \times k/2}$
has rank at most $p+l$.
\end{theorem}
A proof is given in Theorem 1.43 in \cite{Iarrobino1999} for $(p, k) = (2, 6)$, in Theorem 3.2.1 in \cite{Landsberg_2013} for $(p, k, l) = (3, 4, 1)$, and in \S 8 in \cite{Clebsch} for  $(p, k, l) = (3, 4, 2)$.
Finally, for higher $p$ and $k=5$, similarly to the case without latent confounding, we define the matrix $Y_5(T)$ of size  ${\multichoose{p}{2}\binom{p}{a} \times \multichoose{p}{2}\binom{p}{a+1}}$ by
\begin{multline*}
\left(Y_5(T)\right)_{(i_1, \ldots, i_{a+2}), (j_1, \ldots, j_{a+3})} =\\
\begin{cases}
t_{j_1, j_2, i_1, i_2, c} \cdot \text{sgn}\left(\left(
\begin{smallmatrix}
j_3 & \ldots & j_{a+2} & j_{a+3} \\
i_3 & \ldots & i_{a+2} & c
\end{smallmatrix}
\right)\right) & \text{if } \{i_1, \ldots, i_{a+2}\} \subseteq \{j_1, \ldots, j_{a+3}\}, \\
0 & \text{otherwise},
\end{cases}
\end{multline*} where $i_1 \leq i_2$,  $i_3 < \cdots < i_{a+2}$, $j_1 \leq j_2$, $j_3 < \cdots < j_{a+3}$, and $c$ fulfills $\{i_3, \ldots, i_{a+2}, c\} = \{j_3, \ldots ,j_{a+3}\}$.
From Theorem 1.2.3 in \citet{Landsberg_2013}, this matrix drops rank.
\begin{theorem}\label{thm:Y_5} If $T \in \text{Sym}_5\left({\mathbb{R}^p}\right)$ has  symmetric border rank  at most rank $p+l$, then $Y_5(T)$ has at most rank $\binom{p-1}{a}(p+l)$.
\end{theorem}
\subsection{Sufficient Conditions}
Our above exposition highlights testable conditions on moments that hold as a
consequence of the linearity of the structural equations
in~\eqref{eq:introduction_LiNAGAM} and \eqref{eq:introduction_latent_conf}.
The question whether resulting statistical tests can be expected to
have power under alternatives is tied to the question whether tested
constraints are also sufficient for membership in the linear
structural equation model.
The following theorem, which is focused on the case without latent
confounding ($l=0$), shows that sufficiency holds if one mild
additional assumption is added.

\begin{theorem}
  \label{backwards dir}
  Suppose $C^{(2)}$ and $C^{(3)}$ are such that the matrix $M^{(2,3)}$
  has rank at most $p$ and the third moment tensor $C^{(3)}$ has
  symmetric tensor rank at most $p$.  If the matrix $A$ giving a
  representation of $C^{(3)}$ as in~\eqref{def:sym rank} can be be
  chosen to be an invertible $p\times p$ matrix, then $(C^{(2)}, C^{(3)})$ belongs to the
  cumulant model $\mathcal{C}^{(2,3)}_0$.
\end{theorem}
The proof can be found in the supplementary material. For $l>0$,
 Theorem \ref{thm:rank_cond_M}  might be insufficient in terms of precisely characterizing the model $\mathcal{C}^{(k_1, \dots, k_2)}_l$. For $p=2, l=1$, the Theorem still cuts out a set of the correct dimension. Specifically, the dimension of $\mathcal{C}^{(3,4,5)}_1$
matches the dimension of all cumulants $(C^{(3)}, C^{(4)},
C^{(5)})$ satisfying $\text{rank}(M^{(3, 4, 5)})\leq 3$, both being 12.  In contrast, for higher $p$ and $l\geq 1$, Theorem \ref{thm:rank_cond_M} alone does not yield a set of matching dimension, but Theorems \ref{thm:rank_cond_M} and \ref{thm:tensor rank} combined might achieve this. However, in practice, for high $p$, we rely solely on Theorem \ref{thm:rank_cond_M} due to computational efficiency, and, consequently, our test is confined to necessary conditions. That said, we would like to stress that the subset of cumulants
$(C^{(k_1)}, \dots, C^{(k_2)})$ for which $M^{(k_1, \dots, k_2)}$
drops in rank has Lebesgue measure 0 and, thus, a ``randomly'' chosen
point in the alternative, i.e., not conforming with the hypothesized
linear structural equation model, can be expected to
not yield a dropping rank.

\section{Practical Test}
\subsection{Conditions to assess} 
\renewcommand{\arraystretch}{1.5}
\begin{table}[t]
\tbl{Conditions assessed for goodness-of-fit.}{
\begin{tabular}[h]{ccc}
Parameters & Condition stemming from Theorem \ref{thm:rank_cond_M} & Condition stemming from Theorem \ref{thm:tensor rank}\\[3pt]
$l=0$ and $p=2$ & rank$(M^{(2,3)}) \leq p$  & Str$(C^{(3)}) \leq 0$ \\
$l=0$ and $p=3$ & rank$(M^{(2,3)}) \leq p$  & Ar$(C^{(3)}) = 0$  \\
$l=0$ and $p \geq 4$ &  rank$(M^{(2,3)}) \leq p$  & rank$\left(Y_3({C^{(3)}})\right) \leq \binom{p-1}{a}p$\\ 
$l=1$ and $p=2$ & rank$(M^{(3,4,5)}) \leq p+l$ & rank$\left(\text{fl}_3(C^{(6)})\right) \leq p+l$ \\ 
$l\geq 1$ and $p =3$ & rank$(M^{(2,3,4)}) \leq p+l$ & rank$\left(\text{fl}_2(C^{(4)})\right) \leq p+l$ \\
$l\geq 1$ and $p\geq 4$ & rank$(M^{(2,3,4)}) \leq p+l$ & rank$\left(Y_5({C^{(5)}})\right) \leq \binom{p-1}{a}(p+l)$\\
\end{tabular}
}\label{table:Conditions}
\end{table}
\renewcommand{\arraystretch}{1}

In this section, we derive practical tests for the linearity
assumption. Based on the previous section, we assess the conditions
laid out in Table~\ref{table:Conditions}. In principle, all conditions
amount to testing polynomial constraints since rank bounds are
equivalent to the vanishing of certain minors. However, rank constraints may also, and computationally more conveniently, be
assessed via singular value decomposition.  We thus consider different
options for tests. Here, we describe their main aspects relevant for
our setting, while further details are given in the supplement.

\subsection{Characteristic Root Test of Robin and Smith}

Let $\Pi \in \mathbb{R}^{k \times m}$, $k\le m$, be a parameter matrix of a
statistical model.  The characteristic root statistic, abbreviated CR
statistic, of \cite{Robin_2000} is formed to test the null hypothesis
\begin{align*}
  H_0: \: \text{rank}(\Pi) = r\quad\text{vs.}\quad
  H_1: \: \text{rank}(\Pi) > r.
\end{align*}
Let
$\hat\Pi$ be an asymptotically normal estimator of $\Pi$.  Then, the CR
statistic is the sum of the $k-r$ smallest singular values of
$\hat\Pi$, multiplied by the sample size.  The null distribution of
the CR statistic may be asymptotically approximated by a weighted sum
of independent $\chi^2_1$ random variables, the weights being
determined by the asymptotic covariance matrix of $\hat\Pi$. The supplementary material includes further details on the null distribution on local power.

When compared to related methods \citep{Al_Sadoon}, the CR 
statistic has the advantage that the asymptotic approximation also holds if the
asymptotic covariance matrix of $\hat\Pi$ is singular.  Indeed, in our context, the
matrices have duplicated entries leading to duplicated
rows in the asymptotic covariance matrices.

A more subtle issue that remains is that the asymptotic approximation to the
null distribution is justified for $\text{rank}(\Pi) = r$ but not
necessarily for smaller rank \citep{drton:2009}.  
One way this issue can arise in our problem is from Gaussianity.
If at least one component of $\eta$ is Gaussian, the rank of
$M^{(k_1,\ldots,k_2)}$ is strictly lower than $p+l$.  However, in our
simulation studies, the CR test still controls the type I error for
strictly lower rank. Alternatively,  applying the test sequentially for $r=1, \dots, p+l$ is an approach proven to be weakly consistent for the null hypothesis of rank at most $p+l$ \citep[Theorem 5.2]{Robin_2000}.

\subsection{Incomplete U-statistic}
\label{subsec:polytest}
For low dimension $p$, we derived explicit polynomials in the moments. To test these constraints, we consider the methodology of \cite{Sturma},
which utilizes incomplete U-statistics \citep{Chen_2017}.
To briefly describe the methodology, suppose we wish to test whether a collection of polynomials in moments is nonpositive. 
For each individual polynomial $f$, let $h(x_1, \ldots, x_{\deg(f)})$
be a kernel that unbiasedly estimates $f$. Averaging $h$ over independent data points yields the U-statistic
\begin{align}
\label{def u stat}
  U_{n}(f)= 
\frac{1}{\binom{n}{\deg(f)}}
  \sum_{1\leq i_1 < \cdots < i_{\deg(f)} \leq n}  h\left(X^{(i_1)}, \ldots, X^{(i_{\deg(f)})}\right).
\end{align}
We may then test the hypothesis via the maximum of the U-statistics
for the different polynomials.  Its distribution can be approximated via an efficient multiplier bootstrap method.  In this
framework, computational effort may be reduced by using incomplete
U-statistics, which select a random subset of summands in \eqref{def u
  stat}.  As shown in \cite{Sturma}, this also guards against issues
with degenerate kernels, which in our setting may again arise
from $\eta$ having Gaussian cumulants.

\subsection{Combining the Tests}

To obtain an overall test for the linearity assumption, we explore
three approaches:
\begin{itemize}
\item[(i)] Employ the CR statistic to assess the rank of
  $M^{(k_1,\ldots,k_2)}$ but ignore any further tensor
rank constraints.  This test has the same
structure for varying $p$ and $l$ and is computationally fast.
\item[(ii)] Additionally, consider the constraints based on tensor rank, which might lead to improved power. Here, the
incomplete U-statistic is applied if
the constraint takes the form of a polynomial constraint and the CR
statistic is used for a rank constraint. In all cases, the two
results are combined using Bonferroni correction, meaning that for a
given level $\alpha$, we reject if one of the individual tests rejects
at level $\alpha/2$.

  \item[(iii)]  Finally, for $l=0$ and $p\in\{2,3\}$, 
the incomplete U-statistic can be used for both conditions by expressing the
rank constraint in terms of vanishing minors. This
has the advantage that both constraints can be assessed in a unified
manner and no Bonferroni correction is necessary.

\end{itemize}

\section{Simulation Study}\label{sec:simulations}
\subsection{Setup}

We investigate the behaviour of the three proposed approaches and compare them
with the dCovICA method by \cite{Matteson_2017b}. We consider dimension $p\in\{2,4,20\}$ and
$l\in\{0,1\}$ latent variables.  We fix the sample size to $n=1000$ for
$p\in\{2, 4\}$ , to $n=8000$ for $p=20$, and always perform 1000
replications in simulations. Concerning parameter choices arising from the methods,
for the incomplete U-statistic, we use $2n$ as the computational budget
parameter, for dCovICA, we use the asymmetric version, and in both
algorithms we execute 1000 bootstrap replications. Our code can be obtained from \url{https://github.com/DanielaSchkoda/TestLinearSEM}.
For the null hypothesis, we generate data as  
\begin{equation}\tag{H0}\label{eq:H0}
 X  = (I - \Lambda)^{-T} \begin{pmatrix}
  \Gamma^T & I_p
\end{pmatrix} \begin{pmatrix}
L \\ \varepsilon
\end{pmatrix}.
\end{equation}
The exogenous sources $\varepsilon$ and $L$ are drawn independently
from a Gamma distribution with shape and rate parameters drawn
uniformly from  $[2, 3]$ and $[1,5]$, respectively.  The entries of $B$ and $\Gamma$ are chosen at random from $[-1, 1]$.
 
Turning to alternatives, we first consider a violation of the
linearity assumption. More precisely, we simulate $X$ as in \eqref{eq:H0} and
then transform each entry in $X$ by applying the function
\begin{equation}\tag{A1}\label{eq:A1}
f(x) = (1 - \delta) x + \delta  \cos \left(x\right)
\end{equation}
with $\delta$ ranging between $0$ and $1$.  As a second alternative, we sample the data with one more latent confounder than we test for. Specifically, while assessing the null hypothesis with $l$ latent confounders, we generate the data as
 \begin{equation}\tag{A2}\label{eq:A2}
 X  = (I - \Lambda)^{-T} \begin{pmatrix}
   \Gamma^T & I_p
\end{pmatrix} \begin{pmatrix}
L \\ \varepsilon
\end{pmatrix},
 \end{equation}
where $L$ has length $l+1$ and its last component is scaled by $\delta$ varying between $0$ and $5$. 
 We start with the results for $l=0$.

 \subsection{Results}

\begin{figure}[t]
\centering
\begin{subfigure}[c]{0.32\linewidth}
\subcaption{$p=2$}
\includegraphics[width=\linewidth]{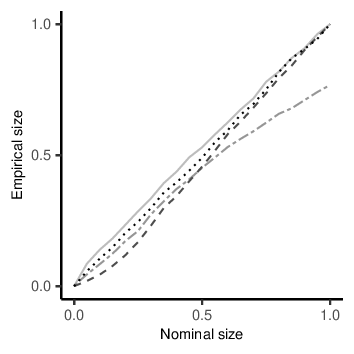}
\end{subfigure}
\begin{subfigure}[c]{0.32\linewidth}
\subcaption{$p=4$}
\includegraphics[width=\linewidth]{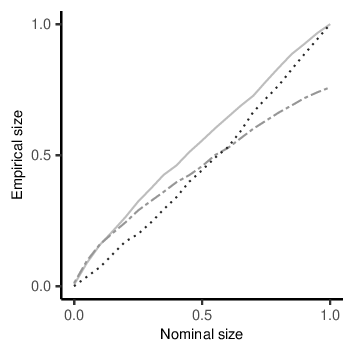}
\end{subfigure}
\begin{subfigure}[c]{0.32\linewidth}
\subcaption{$p=20$}
\includegraphics[width=\linewidth]{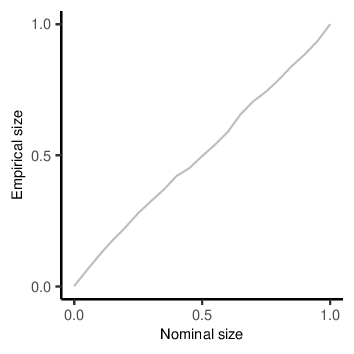}
\end{subfigure}\\
\begin{subfigure}[c]{0.8\linewidth}
\includegraphics[width=\linewidth]{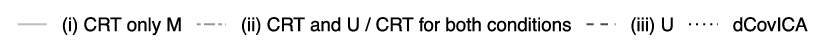}
\end{subfigure}
\caption{Empirical sizes under $H_0$. CRT denotes the CR statistic, and U the incomplete U-statistic.}
\label{fig:H_O}
\end{figure}

\begin{figure}[t]
\centering
\begin{subfigure}[c]{0.32\linewidth}
\subcaption{$p=2$}
\includegraphics[width=\linewidth]{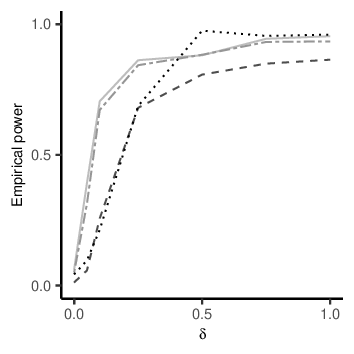}
\end{subfigure}
\begin{subfigure}[c]{0.32\linewidth}
\subcaption{$p=4$}
\includegraphics[width=\linewidth]{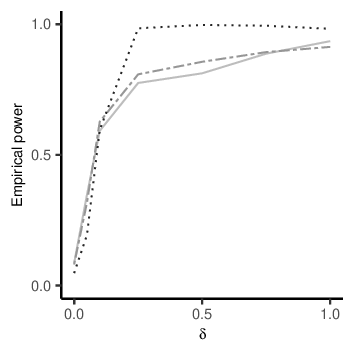}
\end{subfigure}
\begin{subfigure}[c]{0.32\linewidth}
\subcaption{$p=20$}
\includegraphics[width=\linewidth]{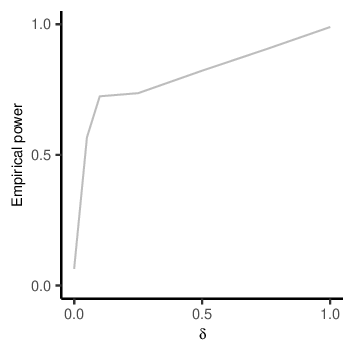}
\end{subfigure}\\
\begin{subfigure}[c]{0.8\linewidth}
\includegraphics[width=\linewidth]{legend.eps}
\end{subfigure}
\caption{Empirical power  against \eqref{eq:A1}.}\label{fig:A1}
\end{figure}

Figure \ref{fig:H_O} shows that under the null hypothesis, the
incomplete U-statistic, as well as dCovICA and the CR statistic hold
the desired level the best. The combinations of the CR statistic with
another procedure have too low sizes for high nominal levels due to
the Bonferroni correction. However, for small nominal levels, they also
perform well.

Turning to the first alternative, as displayed in Fig. \ref{fig:A1},
the power functions of all tests increase quite quickly until they reach values between 0.65 and 1 for $\delta = 1$. The procedures employing the CR statistic exhibit the highest power across almost all values of $\delta$. 

\begin{figure}[t]
\centering
\begin{subfigure}[c]{0.32\linewidth}
\subcaption{$n=1000$}
\includegraphics[width=\linewidth]{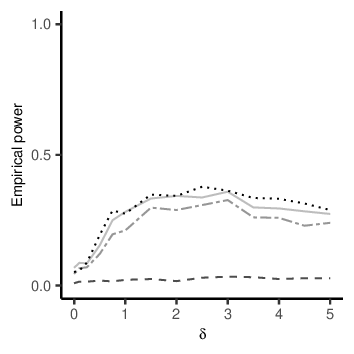}
\end{subfigure}
\begin{subfigure}[c]{0.32\linewidth}
\subcaption{$n=2000$}
\includegraphics[width=\linewidth]{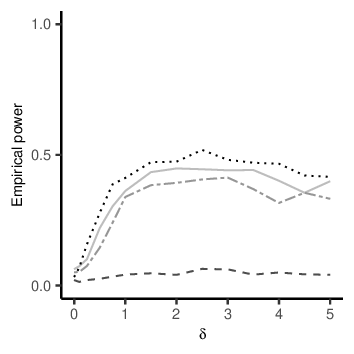}
\end{subfigure}
\begin{subfigure}[c]{0.32\linewidth}
\subcaption{$n=8000$}
\includegraphics[width=\linewidth]{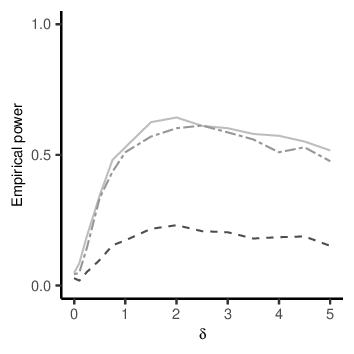}
\end{subfigure}\\
\begin{subfigure}[c]{0.8\linewidth}
\includegraphics[width=\linewidth]{legend.eps}
\end{subfigure}
\caption{Empirical power against \eqref{eq:A2}} \label{fig:A2}
\end{figure}

Figure \ref{fig:A2} depicts the power against alternative
\eqref{eq:A2} for different sample sizes. While the power increases
with higher sample size, it decreases for higher $\delta$, which we
believe to be caused by the additional confounder dominating the noise
terms for high $\delta$ such that the data-generating distribution is
closer to the setting with only one
exogenous source left, which is the confounder.

For $l=1$ and $p=2$, as visualized in Fig.~\ref{fig:latent_conf}, the results are similar to those for $l=0$. In the supplementary material, we include simulation results for other choices of $p$ and $l$.

\begin{figure}[t]
\centering
\begin{subfigure}[c]{0.32\linewidth}
\subcaption{Null hypothesis}
\includegraphics[width=\linewidth]{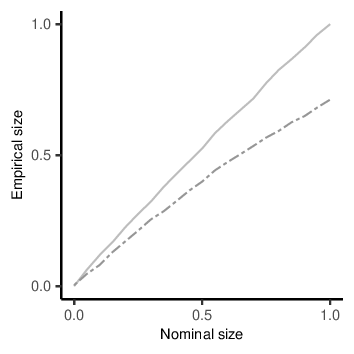}
\end{subfigure}
\begin{subfigure}[c]{0.32\linewidth}
\subcaption{Alternative \eqref{eq:A1}}
\includegraphics[width=\linewidth]{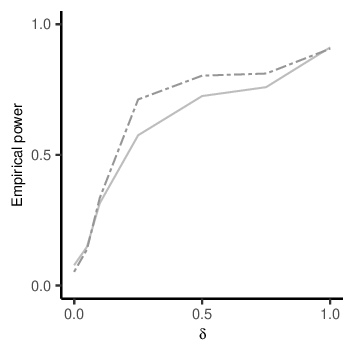}
\end{subfigure}
\begin{subfigure}[c]{0.32\linewidth}
\subcaption{Alternative \eqref{eq:A2}}
\includegraphics[width=\linewidth]{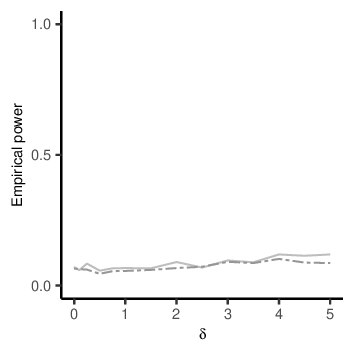}
\end{subfigure}\\
\begin{subfigure}[c]{0.8\linewidth}
\includegraphics[width=\linewidth]{legend.eps}
\end{subfigure}
\caption{Empirical sizes under the null hypothesis and empirical powers against alternatives for $p=2$ and $l=1$.}\label{fig:latent_conf}
\end{figure}

An advantage of our method is the significantly faster computation
time. Table \ref{table:Computing times} shows our benchmarking results on a compute server; the R package
\texttt{microbenchmark} was used for the purpose. For example, for
$p=2$, the CR procedure  is 20000 times faster than dCovICA.

\begin{table}[t]
\tbl{Computation times in milliseconds of each procedure for $l=0$. The average over 1000 repetitions is taken.}{
\begin{tabular}{lcccc}
\multicolumn{1}{c}{} & \multicolumn{1}{c}{CRT only M} & \multicolumn{1}{c}{CRT and U / CRT for both conditions} & \multicolumn{1}{c}{U} &  \multicolumn{1}{c}{dCovICA} \\[5pt]
$p=2$ & \phantom{78626}8 & 10602  & 22452 &  \phantom{1}158105 \\
$p=4$ & \phantom{786}107 & \phantom{10}595 & - &  {1866935} \\
$p=20$ & 786264 &  - & - & -  \\
\end{tabular}}
\label{table:Computing times}
\end{table}
\section{Applications}
\subsection{T\"ubingen Cause-Effect-Pairs} 

\begin{figure}[t]
\centering\includegraphics[width=0.98\linewidth]{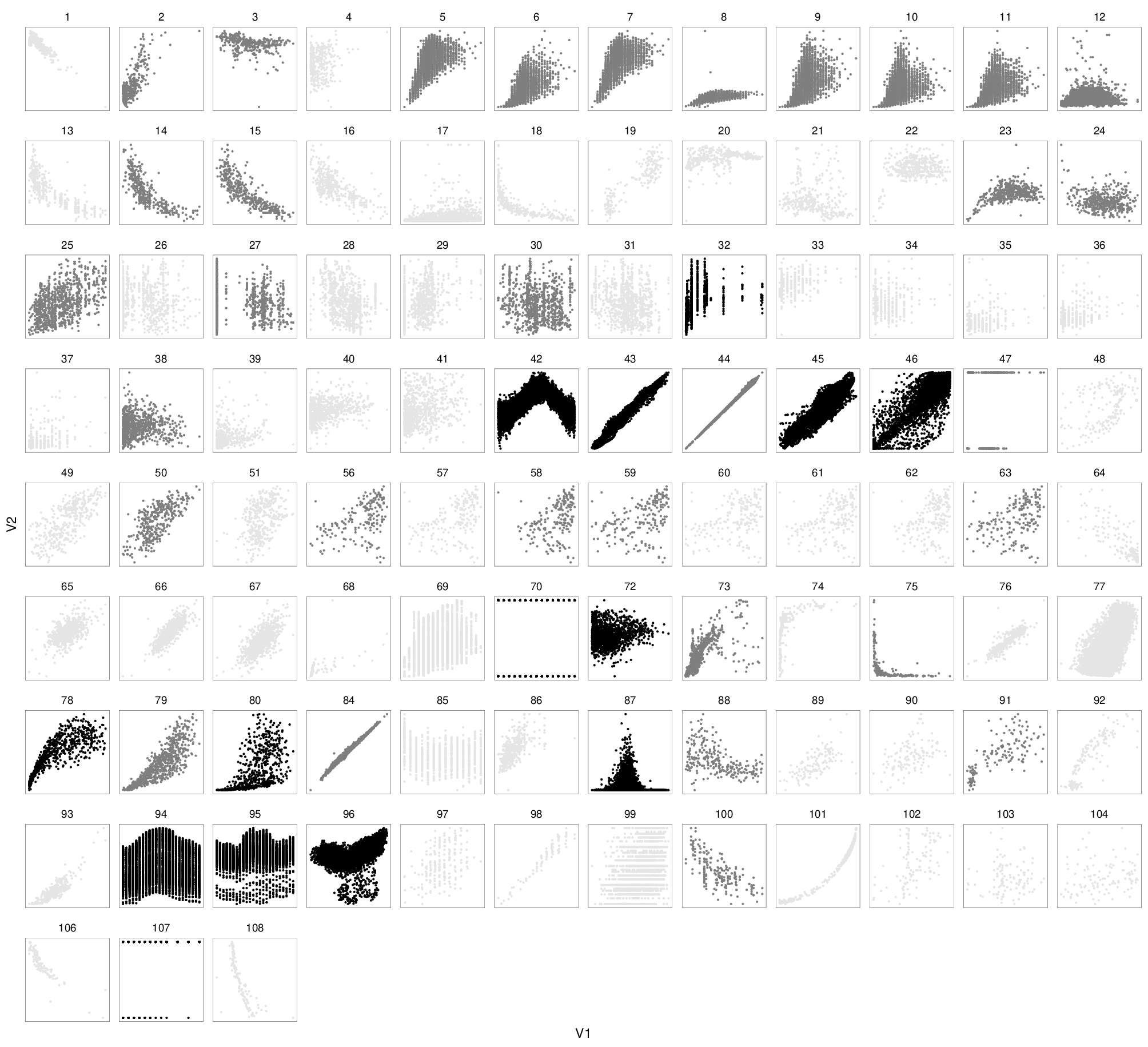}
\caption{Scatter plots of the T\"ubingen pairs. The test for $l=0$ accepted all pairs coloured in light grey at level $\alpha =0.05$, the test for $l=1$ accepted all pairs coloured in dark grey, and both tests rejected the pairs coloured in black.}\label{fig:Tuebingen}
\end{figure}

The T{\"u}bingen cause-effect pairs constitute
a real-world data set collection, which is often used as benchmark to assess the performance of causal inference algorithms \citep{TuebingenPairs}.
It comprises 108 predominantly bivariate data sets, each containing a cause-effect pair with known ground truth. The data stems from different domains, including meteorology, biology, and economics. For example, one of the pairs consists of the hourly wage compared to the age from a study conducted in 1994 and 1995.
We focus on the 99  bivariate data sets and test the goodness-of-fit of
the linear model with $l=0$ as well as  the model with $l=1$ using
procedure (ii). The results are summarized in Fig.
\ref{fig:Tuebingen} and offer a classification of the cause-effect
pairs into a group for which a linear model without confounding is
tenable, a group for which a linear model is tenable after inclusion
of a single confounder, and a group that may be best analysed using
non-linear methods. The supplementary material contains the results for all other methods.

\subsection{Multivariate Data}
As an example of a higher-dimensional problem, we analyse the ecology data from \cite{Grace_2016}, which was collected to evaluate a hypothesis on the relation between ecosystem productivity and plant species richness. 
\begin{figure}[t]
\centering
\includegraphics[width=0.58\linewidth]{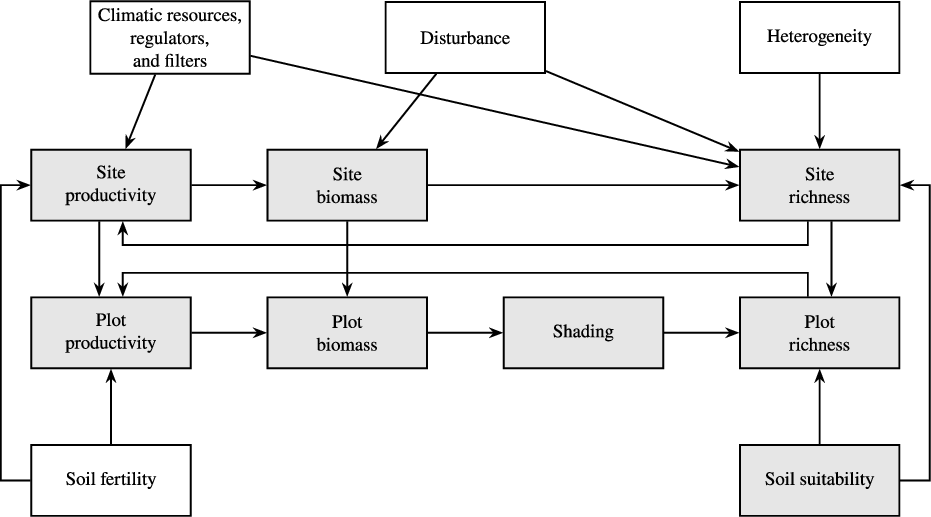}
\caption{Graph describing the causal structure from \cite{Grace_2016}.}\label{fig:SEM_Grace_et_al}
\end{figure}
The initial hypothesis postulates that the underlying mechanisms can
be described by the graph depicted in Fig. 
\ref{fig:SEM_Grace_et_al}, where an edge represents a direct causal
effect. One step in their evaluation is the fitting of a linear
structural equation model with all the variables in the two middle
rows and the variable soil suitability, where they allow for
latent confounding between the variables plot productivity
and plot biomass. Applying our method to this selection of
variables yields a $p$-value 0.003 for $l=0$, indicating a poor fit
of a linear model.  The $p$-value improves to 0.013 for $l=1$, which
is still low, but points to the value of including latent confounders
in model specifications for these data.

\section{Discussion}

Our work provides new goodness-of-fit tests for linear structural
equation models.  Our approach is based on rank constraints that arise
from the algebraic structure of the cumulants. We propose three
variants to test these conditions statistically. While our simulations
suggest that all of them hold level and have state-of-the-art power,
the first suggestion to only assess a matrix rank condition is
computationally favourable, making the method applicable for data sets
with say $p=30$ variables, which is well beyond what can be achieved with
other existing methods.

\section*{Acknowledgment}
This project has received funding from the European Research Council
(ERC) under the European Union’s Horizon 2020 research and innovation
programme (grant agreement No 883818). Daniela Schkoda acknowledges
support by the DAAD programme Konrad Zuse Schools of Excellence in
Artificial Intelligence, sponsored by the Federal Ministry of
Education and Research.

\bibliographystyle{biometrika}
\bibliography{paper-ref}

@book{HyvarinenKO01,
  author       = {Aapo Hyv{\"{a}}rinen and
                  Juha Karhunen and
                  Erkki Oja},
  title        = {Independent Component Analysis},
  publisher    = {Wiley},
  year         = {2001},
  url          = {https://doi.org/10.1002/0471221317},
  doi          = {10.1002/0471221317},
  isbn         = {0-471-40540-X},
  bibsource    = {dblp computer science bibliography, https://dblp.org}
}

@article {drton:2009,
    AUTHOR = {Drton, Mathias},
     TITLE = {Likelihood ratio tests and singularities},
   JOURNAL = {Ann. Statist.},
  FJOURNAL = {The Annals of Statistics},
    VOLUME = {37},
      YEAR = {2009},
    NUMBER = {2},
     PAGES = {979--1012},
      ISSN = {0090-5364},
   MRCLASS = {62H15 (60E05 62H10)},
  MRNUMBER = {2502658},
MRREVIEWER = {Pranesh Kumar},
       DOI = {10.1214/07-AOS571},
       URL = {https://doi.org/10.1214/07-AOS571},
}

@article {Al-Sadoon2017,
    AUTHOR = {Al-Sadoon, Majid M.},
     TITLE = {A unifying theory of tests of rank},
   JOURNAL = {J. Econometrics},
  FJOURNAL = {Journal of Econometrics},
    VOLUME = {199},
      YEAR = {2017},
    NUMBER = {1},
     PAGES = {49--62},
      ISSN = {0304-4076,1872-6895},
   MRCLASS = {62H15 (62H12)},
  MRNUMBER = {3652404},
       DOI = {10.1016/j.jeconom.2017.03.002},
       URL = {https://doi.org/10.1016/j.jeconom.2017.03.002},
}

@article{Mathias_Sam_JMLR,
  author  = {Y. Samuel Wang and Mathias Drton},
  title   = {Causal Discovery with Unobserved Confounding and Non-{G}aussian Data},
   JOURNAL = {J. Mach. Learn. Res.},
  FJOURNAL = {Journal of Machine Learning Research (JMLR)},
  year    = {2023},
  volume  = {24},
  number  = {271},
  pages   = {1--61},
  url     = {http://jmlr.org/papers/v24/21-1329.html}
}

@article {salehkaleybar:2020,
    AUTHOR = {Salehkaleybar, Saber and Ghassami, AmirEmad and Kiyavash,
              Negar and Zhang, Kun},
     TITLE = {Learning linear non-{G}aussian causal models in the presence
              of latent variables},
   JOURNAL = {J. Mach. Learn. Res.},
  FJOURNAL = {Journal of Machine Learning Research (JMLR)},
    VOLUME = {21},
      YEAR = {2020},
     PAGES = {Paper No. 39, 24},
      ISSN = {1532-4435,1533-7928},
   MRCLASS = {62D20},
  MRNUMBER = {4073772},
}

@book {Iarrobino1999,
    AUTHOR = {Iarrobino, Anthony and Kanev, Vassil},
     TITLE = {Power sums, {G}orenstein algebras, and determinantal loci},
    SERIES = {Lecture Notes in Mathematics},
    VOLUME = {1721},
      NOTE = {Appendix C by Iarrobino and Steven L. Kleiman},
 PUBLISHER = {Springer-Verlag, Berlin},
      YEAR = {1999},
     PAGES = {xxxii+345},
      ISBN = {3-540-66766-0},
   MRCLASS = {14M12 (13C40 13H10 13N10 14C05)},
  MRNUMBER = {1735271},
MRREVIEWER = {Juan\ C.\ Migliore},
       DOI = {10.1007/BFb0093426},
       URL = {https://doi.org/10.1007/BFb0093426},
}

@article{schultheiss:2023,
    author = {Schultheiss, C and Bühlmann, P},
    title = "{Ancestor regression in linear structural equation models}",
    journal = {Biometrika},
    pages = {asad008},
    year = {2023},
    month = {03},
    issn = {1464-3510},
    doi = {10.1093/biomet/asad008},
    url = {https://doi.org/10.1093/biomet/asad008},
    eprint = {https://academic.oup.com/biomet/advance-article-pdf/doi/10.1093/biomet/asad008/50570289/asad008.pdf},
}

@article {Comon_2012,
    AUTHOR = {Comon, P. and Ottaviani, G.},
     TITLE = {On the typical rank of real binary forms},
   JOURNAL = {Linear Multilinear Algebra},
  FJOURNAL = {Linear and Multilinear Algebra},
    VOLUME = {60},
      YEAR = {2012},
    NUMBER = {6},
     PAGES = {657--667},
      ISSN = {0308-1087,1563-5139},
   MRCLASS = {15A72 (15A69)},
  MRNUMBER = {2929176},
MRREVIEWER = {Hans\ Arwed\ Keller},
       DOI = {10.1080/03081087.2011.624097},
       URL = {https://doi.org/10.1080/03081087.2011.624097},
}

@article {sen:sen:2014,
    AUTHOR = {Sen, A. and Sen, B.},
     TITLE = {Testing independence and goodness-of-fit in linear models},
   JOURNAL = {Biometrika},
  FJOURNAL = {Biometrika},
    VOLUME = {101},
      YEAR = {2014},
    NUMBER = {4},
     PAGES = {927--942},
      ISSN = {0006-3444,1464-3510},
   MRCLASS = {62J05 (62E20 62F03 62G09 62G10)},
  MRNUMBER = {3286926},
       DOI = {10.1093/biomet/asu026},
       URL = {https://doi.org/10.1093/biomet/asu026},
}

@article {heinze:2018,
    AUTHOR = {Heinze-Deml, Christina and Maathuis, Marloes H. and
              Meinshausen, Nicolai},
     TITLE = {Causal structure learning},
   JOURNAL = {Annu. Rev. Stat. Appl.},
  FJOURNAL = {Annual Review of Statistics and its Application},
    VOLUME = {5},
      YEAR = {2018},
     PAGES = {371--394},
      ISSN = {2326-8298,2326-831X},
   MRCLASS = {62H99 (94A08 94C30)},
  MRNUMBER = {3774752},
       DOI = {10.1146/annurev-statistics-031017-100630},
       URL = {https://doi.org/10.1146/annurev-statistics-031017-100630},
}

@incollection {handbook:spirtes,
    AUTHOR = {Spirtes, Peter and Zhang, Kun},
     TITLE = {Search for causal models},
 BOOKTITLE = {Handbook of graphical models},
    SERIES = {Chapman \& Hall/CRC Handb. Mod. Stat. Methods},
     PAGES = {439--469},
 PUBLISHER = {CRC Press, Boca Raton, FL},
      YEAR = {2019},
      ISBN = {978-1-4987-8862-5},
   MRCLASS = {62H99 (62A01 62J02 62J05)},
  MRNUMBER = {3888007},
}

@article{DirectLiNGAM,
  title={{DirectLiNGAM}: A direct method for learning a linear non-{G}aussian structural equation model},
  author={Shimizu, S. and Inazumi, T. and Sogawa, Y. and Hyv{\"a}rinen, A. and Kawahara, Y. and Washio, T. and Hoyer, P. O. and Bollen, K.},
  journal={J.  Mach.  Learn.  Res.},
  volume={12},
  pages={1225--1248},
  year={2011}
}

@article{ICALiNGAM,
  title={A linear non-{G}aussian acyclic model for causal discovery.},
  author={Shimizu, S. and Hoyer, P. O. and Hyv{\"a}rinen, A. and Kerminen, A. and Jordan, M.},
  journal={J.  Mach.  Learn.  Res.},
  volume={7},
  number={10},
  year={2006},
  pages={2003--2030}
}

@book{McCullagh_2018,
 author = {McCullagh, P.},
 title = {Tensor methods in statistics},
 edition={second},
  year      = 2018,
  publisher = "Dover Publications",
}

@book{Comon_2010,
author = {P. Comon and C. Jutten},
title = {Handbook of Blind Source Separation},
publisher = {Academic Press},
address = {Oxford},
pages = {107--154},
year = {2010}
}

@article{Hoyer_2008,
 author = {Hoyer, P. O. and Shimizu, S. and Kerminen, A. J. and Palviainen, M.},
 year = {2008},
 title = {Estimation of causal effects using linear non-{G}aussian causal models with hidden variables},
 url = {https://www.sciencedirect.com/science/article/pii/S0888613X08000212},
 pages = {362--378},
 volume = {49},
 number = {2},
 issn = {0888613X},
 journal = {Internat. J. Approx. Reason.},
 doi = {10.1016/j.ijar.2008.02.006}
}

@book{book_shimizu,
  title     = "Statistical causal discovery: {LiNGAM} approach",
  author    = "Shimizu, S.",
  year      = 2022,
  publisher = "Springer Tokyo",
  doi = "https://doi.org/10.1007/978-4-431-55784-5"
}

@article {Mathias_2020,
    AUTHOR = {Wang, Y. S. and Drton, M.},
     TITLE = {High-dimensional causal discovery under non-{G}aussianity},
   JOURNAL = {Biometrika},
    VOLUME = {107},
      YEAR = {2020},
    NUMBER = {1},
     PAGES = {41--59},
   MRCLASS = {62D20 (05C90 62A09 62H22)},
  MRNUMBER = {4064139},
MRREVIEWER = {Nicholas T. Longford},
       DOI = {10.1093/biomet/asz055}
}

@book {Handbook,
     TITLE = {Handbook of graphical models},
    SERIES = {Chapman \& Hall/CRC Handbooks of Modern Statistical Methods},
    EDITOR = {Maathuis, M. and Drton, M. and Lauritzen, S.
              and Wainwright, M.},
 PUBLISHER = {CRC Press, Boca Raton, FL},
      YEAR = {2019},
     PAGES = {xviii+536},
   MRCLASS = {62-00 (62-06 62H99)},
  MRNUMBER = {3889064},
}

@InProceedings{Raghu_2018,
  title = 	 {Evaluation of Causal Structure Learning Methods on Mixed Data Types},
  author = 	 {Raghu, Vineet K. and Poon, Allen and Benos, Panayiotis V.},
  booktitle = 	 {Proceedings of 2018 ACM SIGKDD Workshop on Causal Disocvery},
  pages = 	 {48--65},
  year = 	 {2018},
  opteditor = 	 {Le, Thuc Duy and Zhang, Kun and Kıcıman, Emre and Hyvärinen, Aapo and Liu, Lin},
  volume = 	 {92},
  series = 	 {Proceedings of Machine Learning Research},
  month = 	 {20 Aug},
  publisher =    {PMLR},
  url = 	 {https://proceedings.mlr.press/v92/raghu18a.html},
}

@article{Robin_2000,
 ISSN = {02664666, 14694360},
 URL = {http://www.jstor.org/stable/3533192},
 author = {J. Robin and R. J. Smith},
 journal = {Econometric Theory}, 
 number = {2},
 pages = {151--175},
 publisher = {Cambridge University Press},
 title = {Tests of Rank},
 urldate = {2023-02-17},
 volume = {16},
 year = {2000}
}

@article{aronhold,
  title     = {{Theorie der homogenen Funktionen dritten Grades von drei Ver\"anderlichen}},
  journal     = { J. Reine Angew. Math. },
  author = {S. Aronhold},
  publisher = { de Gruyter },
  year      = { 1858 }, 
  volume   = {55},
  pages = {97--191}}

@article{Ottaviani, 
title={An Invariant Regarding {W}aring’s Problem for Cubic Polynomials},
 volume={193}, 
 DOI={10.1017/S0027763000026040}, 
 journal={Nagoya Math. J.}, 
 publisher={Cambridge University Press}, 
 author={Ottaviani, G.}, 
 year={2009}, 
 pages={95--110}
}

@book{book_landsberg,
  title     = "Tensors: Geometry and applications",
  author    = "J. M. Landsberg",
  year      = 2012,
  publisher = "American Mathematical Society",
  series = "Graduate Studies in Mathematics",
  address = "Providence, Rhode Island",
  volume = 128
}

@Article{apolarity_lemma,
author={Blekherman, G.},
title={Typical Real Ranks of Binary Forms},
journal={Found. Comput. Math.},
year={2015},
month={Jun},
day={01},
volume={15},
number={3},
pages={793--798},
issn={1615-3383},
doi={10.1007/s10208-013-9174-8},
url={https://doi.org/10.1007/s10208-013-9174-8}
}

@article{Al_Sadoon,
title = {A unifying theory of tests of rank},
journal = {J. Econometrics},
volume = {199},
number = {1},
pages = {49--62},
year = {2017},
doi = {https://doi.org/10.1016/j.jeconom.2017.03.002},
author = {M. M. Al-Sadoon}
}

@misc{Sturma,
  doi = {10.48550/ARXIV.2208.11756},
  url = {https://arxiv.org/abs/2208.11756},
  author = {Sturma, N. and Drton, M. and Leung, D.},
  title = {Testing Many and Possibly Singular Polynomial Constraints},  
  year = {2022},
  note = {arXiv preprint}
}

@article{TuebingenPairs,
  author  = {J. M. Mooij and J. Peters and D. Janzing and J. Zscheischler and B. Sch{{\"o}}lkopf},
  title   = {Distinguishing Cause from Effect Using Observational Data: Methods and Benchmarks},
  journal = {J.  Mach.  Learn.  Res.},
  year    = {2016},
  volume  = {17},
  number  = {32},
  pages   = {1--102}
}

@article{Landsberg_2013,
 author = {Landsberg, J. M. and Ottaviani, G.},
 year = {2013},
 title = {Equations for secant varieties of {V}eronese and other varieties},
 url = {https://link.springer.com/article/10.1007/s10231-011-0238-6},
 pages = {569--606},
 volume = {192},
 number = {4},
 issn = {1618-1891},
 journal = {Ann. Mat. Pura Appl.},
 doi = {10.1007/s10231-011-0238-6}
}

@article{MaurizioBanchi_2015,
 author = {M. Banchi},
 year = {2015},
 title = {Rank and border rank of real ternary cubics},
 journal = {Boll. Unione Mat. Ital.},
 volume = {8},
 pages = {65--80}
}

@article {Seigal,
    AUTHOR = {Seigal, A.},
     TITLE = {Ranks and symmetric ranks of cubic surfaces},
   JOURNAL = {J. Symbolic Comput.},
  FJOURNAL = {Journal of Symbolic Computation},
    VOLUME = {101},
      YEAR = {2020},
     PAGES = {304--317},
      ISSN = {0747-7171,1095-855X},
   MRCLASS = {14Q10 (15A03 15A69 51N35)},
  MRNUMBER = {4109719},
MRREVIEWER = {Douglas\ A.\ Torrance},
       DOI = {10.1016/j.jsc.2019.10.001},
       URL = {https://doi.org/10.1016/j.jsc.2019.10.001},
}

@article {Clebsch,
    AUTHOR = {Clebsch, A.},
     TITLE = {Ueber {C}urven vierter {O}rdnung},
   JOURNAL = {J. Reine Angew. Math.},
  FJOURNAL = {Journal f\"{u}r die Reine und Angewandte Mathematik. [Crelle's
              Journal]},
    VOLUME = {59},
      YEAR = {1861},
     PAGES = {125--145},
      ISSN = {0075-4102,1435-5345},
   MRCLASS = {99-04},
  MRNUMBER = {1579172},
       DOI = {10.1515/crll.1861.59.125},
       URL = {https://doi.org/10.1515/crll.1861.59.125},
}

@article{Matteson_2017b,
 author = {Matteson, D. S. and Tsay, R. S.},
 year = {2017},
 title = {Independent Component Analysis via Distance Covariance},
 pages = {623--637},
 volume = {112},
 number = {518},
 issn = {0162-1459},
 journal = {J. Amer. Statist. Assoc.},
 doi = {10.1080/01621459.2016.1150851}
}

@article{Chen_2017,
  title={Randomized incomplete {$U$}-statistics in high dimensions},
  author={X. Chen and K. Kato},
  journal={Ann. Statist.},
  volume={47},
  number={6},
  pages={3127--3156},
  year={2019}
}

@article{Grace_2016,
 author = {Grace, J. B. and Anderson, T. M. and Seabloom, E. W. and Borer, E. T. and Adler, Peter B. and Harpole, W. Stanley and Hautier, Yann and Hillebrand, Helmut and Lind, Eric M. and P{\"a}rtel, Meelis and Bakker, Jonathan D. and Buckley, Yvonne M. and Crawley, Michael J. and Damschen, Ellen I. and Davies, Kendi F. and Fay, Philip A. and Firn, Jennifer and Gruner, Daniel S. and Hector, Andy and Knops, Johannes M. H. and MacDougall, Andrew S. and Melbourne, Brett A. and Morgan, John W. and Orrock, John L. and Prober, Suzanne M. and Smith, Melinda D.},
 year = {2016},
 title = {{Integrative modelling reveals mechanisms linking productivity and plant species richness}},
 pages = {390--393},
 volume = {529},
 number = {7586},
 journal = {{Nature}}
}

@inproceedings{Lacerda_2008,
author = {Lacerda, G. and Spirtes, P. and Ramsey, J. and Hoyer, P. O.},
title = {Discovering Cyclic Causal Models by Independent Components Analysis},
year = {2008},
isbn = {0974903949},
publisher = {AUAI Press},
booktitle = {Proc. of the 24th Conf. on Uncertainty in Artificial Intelligence (UAI)},
pages = {366–-374},
numpages = {9},
location = {Helsinki, Finland},
}

@inproceedings{Strieder_2021,
	author = {Strieder, D. and  Freidling, T. and  Haffner, S. and  Drton, M. },
	title = {Confidence in causal discovery with linear causal models},
	booktitle = {Proc. of the 37th Conf. on Uncertainty in Artificial Intelligence (UAI)},
	year = {2021},
	editor = {de Campos, C. and Maathuis, M. H.},
	volume = {161},
	month = {Jul},
	pages = {1217--1226},
	publisher = {PMLR}
}

@inproceedings{Biza_2020,
  title = 	 {Tuning Causal Discovery Algorithms},
  author =       {Biza, K. and Tsamardinos, I. and Triantafillou, S.},
  booktitle = 	 {Proc. of the 10th International Conf. on Probabilistic Graphical Models},
  pages = 	 {17--28},
  year = 	 {2020},
  volume = 	 {138},
  editor = 	 {Jaeger, M. and Nielsen, T. D.},
publisher =    {PMLR}
}

@inproceedings{Zhang_2009,
author = {Zhang, K. and Hyv\"{a}rinen, A.},
title = {On the Identifiability of the Post-Nonlinear Causal Model},
year = {2009},
publisher = {AUAI Press},
booktitle = {Proc. of the 25th Conf. on Uncertainty in Artificial Intelligence (UAI)},
pages = {647–-655},
location = {Montreal, Quebec, Canada},
}

\clearpage
\begin{center}
{\LARGE \bfseries Supplementary material for ``Goodness-of-fit tests for linear non-Gaussian structural equation models''}
\end{center}
    \vskip0.2in
    This supplement
    includes all omitted proofs as well as additional
numerical experiments. These experiments consider the impact of
Gaussian noise and they report on other choices of the dimension $p$
and the number of latents $l$ to cover all six cases listed in Table~\ref{table:Conditions}.
 
\appendix

\section{Proofs}
\subsection{Proofs for Tensor Rank Conditions}
In this subsection, we give the omitted proofs for Section 3.2. We first introduce some notation and basic concepts concerning tensor spaces. A more detailed introduction can be found in \cite{book_landsberg}. By $\mathbb{K}$ we denote the field of real or complex numbers. Let $V$ be a real or complex finite-dimensional vector space. Then, denote by $V^*$ the dual of $V$, and by 
$v^{\vee} \in V^*$ the associated dual element of an element $v \in
V$.  Let
$\mathfrak{S}_k = \{\pi: [k] \to [k]: \pi \text{ permutation on the
set } [k]\}$ be the symmetric group of order $k$, let
$\text{sgn}(\pi)$ be the sign of an element $\pi \in \mathfrak{S}_k$, and let
$e_i$ be the $i$th standard basis vector of $\mathbb{K}^q$.
The space of tensors $\left(\mathbb{K}^{q}\right)^{\otimes k}$ can be identified with the space of multilinear maps
\begin{equation*}
\{Q: \left(\mathbb{K}^q\right)^k \to \mathbb{K}: Q \text{ is multilinear}\}
\end{equation*}
as follows. To an array $T \in \left(\mathbb{K}^{q}\right)^{\otimes k}$, we associate the multilinear map  given by
\begin{align*}
Q(e_{i_1}, \dots, e_{i_k}) = t_{i_1, \dots \i_k}, \quad (i_1, \dots, i_k \in [q]).
\end{align*}
Conversely, for a multilinear form $Q: \left(\mathbb{K}^q\right)^k   \to \mathbb{K}$ the respective array is defined by
\begin{align*}
t_{i_1, \dots, i_k} = Q(e_{i_1}, \dots, e_{i_k}), \quad (i_1, \dots, i_k \in [q]).
\end{align*}
Furthermore, for $v_1,  \dots, v_k \in \mathbb{K}^q$, we define their tensor product $v_1 \otimes  \dots  \otimes  v_k$ by
\begin{align*}
(v_1 \otimes  \dots  \otimes  v_k)(w_1, \dots, w_k) = \prod_{i=1}^k v_i^{\vee}(w_i), \quad (w_1, \dots, w_k \in \mathbb{K}^q),
\end{align*}
and their wedge product as
\begin{align*}
v_1 \wedge \dots \wedge v_k =  \frac{1}{k!} \sum_{\pi \in \mathfrak{S}_k} \text{sgn}(\pi) v_{\pi(1)} \otimes \cdots \otimes  v_{\pi(k)}.
\end{align*}
The space of symmetric tensors $\text{Sym}_k(\mathbb{K}^q)$ can be identified with the space of homogeneous polynomials on $\mathbb{K}^q$ with degree $k$ as follows. Given a symmetric tensor $Q \in \text{Sym}_k(\mathbb{K}^q)$ viewed as a multilinear form $Q: \left(\mathbb{K}^q\right)^k \to \mathbb{K}$, the corresponding polynomial is 
\begin{align*}
f(x) = Q(x, \dots, x), \quad \left(x \in \mathbb{K}^q\right).
\end{align*}
The inverse map maps a polynomial  $f$ to the multilinear form $Q$ given by
$$Q(x_{1},\dots, x_{k})=\frac{1}{k!}\sum_{I \subseteq [k], I \neq \emptyset}(-1)^{k-|I|}f\left(\sum_{i\in I}x_{i}\right), \quad (x_1, \dots, x_k \in \mathbb{K}^q).$$
In this representation, rank one tensors are elements of the form $\zeta^k$ with $\zeta: \mathbb{K}^q \to \mathbb{K}$ a linear function.

Now, we turn to the proof of Theorem 3. Throughout the whole proof, we
work over the real field. A central tool is the \textit{Apolarity
Lemma}, which was first proven by Sylvester;  our reference is \citet[Lemma 2.1]{apolarity_lemma}. To state it, we need the notion of an \textit{apolar ideal}.
\begin{definition} The apolar ideal $f^\perp$ of a homogeneous polynomial $f \in \mathbb{R}[x, y]$ is the set of all polynomials whose differential operator annihilates $f$
\begin{align*}
f^\perp = \{h \text{ homogeneous polynomial in } \mathbb{R}[x, y]: \partial h (f) = 0\}
\end{align*}
where the differential operator $\partial h$ is defined as
\begin{align*}
\partial h=\sum_{i=0}^{d} \lambda_{i}{\frac{\partial^{d}}{\partial x^{i}\partial y^{d-i}}}.
\end{align*} for $h =\sum_{i=0}^{d} \lambda_{i}x^{i}y^{d-i}$.
\end{definition}
\begin{lemma}[Apolarity Lemma] \label{lemma:apolarity}
Let $f = \sum_{i=1}^3 \alpha_i x^i y^{3-i}\in\mathbb{R}[x, y]$ be a
homogeneous polynomial of degree $3$. Then $f$ can be written as a
linear combination of two rank one tensors, i.e., 
\begin{align*}
f = \delta_1 (\beta_1 x + \gamma_1 y)^3 + \delta _2 (\beta_2 x
+\gamma_2 y)^3
\end{align*} 
for $\beta_1,\beta_2,\gamma_1,\gamma_2\in\mathbb{R}$
if and only if $g = (\gamma_1 x - \beta_1 y)(\gamma_2 x - \beta_2 y)$ is contained in the apolar ideal $f^\perp$.
\end{lemma}
\begin{proof}[ of Theorem 3]
From the apolarity lemma, $f$ has real border rank at most two if and only if there exists a real polynomial $g$ of degree two such that
\begin{enumerate} 
\item[(i)] $g$ is contained in the apolar ideal of $f$, and
\item[(ii)] $g$ can be written as $g = (\gamma_1 x - \beta_1
y)(\gamma_2 x - \beta_2 y)$ for $\beta_1,\beta_2,\gamma_1,\gamma_2\in\mathbb{R}$.
\end{enumerate}
Denote $g = \sum_{i=0}^{2} g_{i}x^{i}y^{2-i}$. Condition (ii) yields the  equation system 
\begin{align*}
g_0 &= \beta_1 \beta_2\\
g_1 &= - \beta_1 \gamma_2 - \beta_2 \gamma_1\\
g_2 &= \gamma_1 \gamma_2,
\end{align*}
which has a real solution precisely if $g_1^2 - 4g_0 g_2 \geq 0$. For condition (i), we calculate 
\begin{align*}
\partial g(f) = \left( 6 \alpha_3  g_2 + 2 \alpha_1  g_1 + 2 \alpha_2  g_2 \right)x + \left(6 \alpha_0  g_0 +2 \alpha_1 g_0 + 2 \alpha_2 g_1\right)y.
\end{align*}
Thus, $\partial g(f) = 0$ is equivalent to 
\begin{align}
\left(\begin{array}{c} g_1 \\ g_2 \\ g_3 \end{array}\right)
 \in \left\lbrace \left(\begin{array}{c} \mu \\ -\frac{\mu \left(9 \alpha _0 \alpha _3-\alpha _1 \alpha _2\right)}{3 \alpha _1 \alpha _3-\alpha _2^2} \\ -\frac{\mu \left(3 \alpha _0 \alpha _2-\alpha _1^2\right)}{\alpha _2^2-3 \alpha _1 \alpha _3} \end{array}\right): \mu \in \mathbb{R}
\right\rbrace
\end{align}
if $3 \alpha _1 \alpha _3-\alpha _2^2 \neq 0$, and
\begin{align*}
\left(\begin{array}{c} g_1 \\ g_2 \\ g_3 \end{array}\right)
 \in \left\lbrace \left(\begin{array}{c} 0 \\ \mu \\ \sqrt{\frac{\alpha_1}{3\alpha_3}} \mu \end{array}\right): \mu \in \mathbb{R}
\right\rbrace
\end{align*}
otherwise. This condition combined with the requirement $g_1^2 - 4g_0 g_2 \geq 0$ yields the inequality appearing in Lemma \ref{lemma:apolarity}.
\end{proof}

\textit{Explicit construction of the Young flattening $Y_k(T)$}. To derive the explicit representation of  $Y_3(T)$ given in Theorem 5, we choose 
\[
\left\{ e_{i_1} \otimes (e_{i_2} \wedge \dots 
\wedge e_{i_{a+1}}): i_1, \dots,i_{a+1} \in [p], \text{ and } i_2 < \dots < i_{a+1} \right\}
\] 
as basis for the domain of definition, and 
\[\left\{e_{j_1} \otimes (e_{j_2} \wedge \dots 
\wedge e_{j_{a+2}}): j_1, \dots,j_{a+2} \in [p],\text{ and } j_2 < \dots < j_{a+2}\right\}\] as basis for the image space. Similarly, to arrive at the matrix formula for $Y_5(T)$ as stated in Theorem 7 we work with 
\[\left\{\frac{1}{2}(e_{i_1} \otimes e_{i_2}+e_{i_2} \otimes e_{i_1} ) \otimes (e_{i_3} \wedge \dots 
\wedge e_{i_{a+2}}): i_1, \dots,i_{a+2} \in [p], i_1 \leq i_2, \text{ and } i_3 < \dots < i_{a+2}\right\}\] as basis for the definition space, and 
\[\left\{\frac{1}{2}(e_{j_1} \otimes e_{j_2}+e_{j_2} \otimes e_{j_1} ) \otimes (e_{j_2} \wedge \dots 
\wedge e_{j_{a+2}}): j_1, \dots,j_{a+3} \in [p], j_1 \leq j_2, \text{ and } j_3 < \dots < j_{a+3}\right\}\] as basis for the image space.

\subsection{Proof of Theorem 8}
\begin{proof} We need to find an invertible matrix $B$ with its inverse having ones on the diagonal, and diagonal tensors $\Omega^{(2)} \in \mathbb{R}^{p \times p}$, $\Omega^{(3)} \in \mathbb{R}^{p \times p \times p}$ such that
\[C^{(2)} = \Omega^{(2)} \bullet B \bullet B, \quad C^{(3)} = \Omega^{(3)} \bullet B \bullet B \bullet B. \]
By assumption, there exists an invertible matrix $A$ fulfilling \[c^{(3)}_{jkl} = \sum_{i=1}^r a_{ji}a_{ki}a_{li}.\] Denoting by $D$ the diagonal matrix with entries $d_{ii} = (A^{-1})_{ii}$, this yields
\begin{equation}\label{eq:expression_C3}
C^{(3)} = \Omega^{(3)} \bullet B \bullet B \bullet B
\end{equation}
for  $B = AD$  and $\Omega^{(3)}$ the diagonal tensor with $\omega^{(3)}_{iii} = d_{ii}^{-3}$. To derive  the expression for $C^{(2)}$, we flatten both sides of equation \eqref{eq:expression_C3}, resulting in
\begin{align*}
fl_2\left(C^{(3)} \right) = 
B 
\begin{pmatrix}
\omega^{(3)}_{111} &    &  \\
  & \ddots &   \\
 &   &  \omega^{(3)}_{ppp} 
\end{pmatrix} 
\begin{pmatrix}
b_{11}b_{11} & b_{11}b_{21} & \cdots & b_{p1}b_{p1} \\
b_{12}b_{12} & b_{12}b_{22} & \cdots & b_{p2}b_{p2} \\
\vdots  & \vdots  & \ddots & \vdots  \\
b_{1p}b_{1p} & b_{1p}b_{2p} & \cdots & b_{pp}b_{pp} 
\end{pmatrix}.
\end{align*}
The first two factors are invertible by their definition. 
To show that the last factor has linearly independent rows, assume that there are $\alpha_i$, $i \in [p]$ such that
\begin{align*}
\alpha_1 b_{j1} b_{k1} + \dots + \alpha_p b_{jp} b_{kp} = 0, \quad (j,
k \in [p], \; j \leq k).
\end{align*}
Setting $j=1$, we derive
\begin{align*}
\left(\alpha_1 b_{11}\right) b_{k1} + \dots + \left(\alpha_p b_{1p}\right) b_{kp} = 0, \quad (k \in [p]).
\end{align*}
The linear independence of the columns of $B$ yields that for each $k\in [p]$, either $\alpha_k$ or $b_{k1}$ is zero. Similarly, one can conclude that for each $j \in [p]$ and for each $k \in [p]$, $\alpha_k$ is zero or $b_{jk}$ is zero. Combined, we obtain for that each $k \in [p]$, $
\alpha_k = 0$ or $b_{jk} = 0$ for all $j \in [p]$.
The second option would contradict the invertibility of $B$. Hence, all $\alpha_k$ are zero, which shows that the last factor, and consequently $fl_2\left(C^{(3)} \right)$ have linearly independent rows.

Since the lower $p$ rows of $M^{(2,3)}$ coincide with $fl_2\left(C^{(3)} \right)$ and $M^{(2,3)}$ has rank $p$, we obtain that the first row of $M^{(2,3)}$ can be written as a linear combination of the rows of $fl_2\left(C^{(3)} \right)$. In particular, there exists a $\zeta \in \mathbb{R}^p$ fulfilling
\begin{align*}
c^{(2)}_{jk} 
= \sum_{\nu=1}^p \zeta_{\nu} c^{(3)}_{jk \nu} 
= \sum_{\nu=1}^p \zeta_{\nu} \sum_{i=1}^p \omega^{(3)}_{iii} b_{ji}b_{ki}b_{\nu i}.
\end{align*}
Therefore, with $\omega^{(2)}_{ii} = \sum_{\nu} \zeta_{\nu} \omega^{(3)}_{iii}b_{ \nu i}$,
\begin{align*}
C^{(2)} =  \Omega^{(2)} \bullet B \bullet B,
\end{align*}
which concludes the proof.
\end{proof}
\section{Background on the Statistical Methods Used}\label{sec:statistical_methods}
\subsection{Characteristic Root Test of Robin and Smith}
Recall that the CR statistic by \cite{Robin_2000}  assesses the null hypothesis
\begin{align*}
  H_0: \: \text{rank}(\Pi) = r\quad\text{vs.}\quad
  H_1: \: \text{rank}(\Pi) > r.
\end{align*}
for a parameter matrix $\Pi \in \mathbb{R}^{k \times m}$, $k \leq
m$. It requires the existence of an asymptotically normal estimator
$\hat{\Pi}$ of $\Pi$, so as the sample size $n$ tends to infinity we have
\begin{equation*}
\sqrt{n} \left(\text{vec}(\Pi) - \text{vec}(\hat{\Pi}) \right)
\to \mathcal{N}(0, W)
\end{equation*}
in distribution for some asymptotic covariance matrix $W$. As the name indicates, the test is based on the singular values of the matrix $\Pi$, which are the roots of the characteristic polynomial of $\Pi^T\Pi$. More precisely, we leverage that a matrix has rank at most $r$ if and only if all singular values starting from the $(r+1)$th are zero. We denote the singular value decomposition of $\Pi$ by
$
\Pi =U \Sigma V^T,
$ where $ U \in \mathbb{R}	^{k \times k}$ and $ V \in \mathbb{R}^{m \times m}$. We write $ \sigma_1, \dots,  \sigma_k$ for the singular values of $\Pi$. Furthermore, we consider the decompositions $
 U = \begin{pmatrix}
 U_1 &   U_2
\end{pmatrix} \text{ and }  V = \begin{pmatrix}
 V_1 &   V_2
\end{pmatrix},
$ with $U_1 \in \mathbb{R}^{k \times r},\:  U_2 \in \mathbb{R}^{k \times r-k},\:  V_1 \in \mathbb{R}^{m \times r}$, and $ V_{2} \in \mathbb{R}^{m \times m-r}$. Similarly, $\hat\Pi =\hat U \hat\Sigma \hat V^T$ denotes the singular value decomposition of $\hat\Pi$.
With this notation, the test statistic is defined as
\begin{align*}
\text{CRT}_n = n \sum_{i=r+1}^k \hat  \sigma_i^2.
\end{align*}
Under the null hypothesis, the statistic's limiting distribution is a weighted sum of chi-square distributions. Specifically,
\begin{align*}
\text{CRT}_n \to \sum_{i=1}^l \lambda_i \chi^2_1
\end{align*} in distribution, where $\lbrace \lambda_i \rbrace_{i=1}^l$ are the non-zero eigenvalues of
$\Omega = \left(V_2^T \otimes U_2^T\right) W \left(V_2 \otimes U_2\right)$ \cite[Theorem 3.2]{Robin_2000}. Hence, a consistent test can be obtained by first retrieving estimates $\hat{\lambda}_1, \dots , \hat{\lambda}_l$ of the eigenvalues $\lambda_1, \dots \lambda_l$ and then rejecting if 
\begin{align*}
\text{CRT}_n \geq q_\alpha,
\end{align*} where $\alpha$ is the nominal level, and $q_\alpha$ is the $\alpha$-quantile of the distribution $\sum_{i=1}^l \hat{\lambda}_i \chi^2_1$, see \citet[Theorem 5.1]{Robin_2000}. Moreover, the test has power against local alternatives, as follows.
\begin{theorem}
\label{thm:local-power}
  Under a local alternative of the form $\Pi = \Psi + \sqrt{n}^{-1}\Theta$, with matrix $\Psi$ of rank $r$, and $\Theta$ of rank $r^* > r$,
\begin{equation*}
    \text{CRT}_n \to \sum_{i=1}^l \lambda_i \chi^2\left(\frac{s_i^T \text{vec}(\Sigma_2)_i^2}{\lambda_i}, 1\right)
\end{equation*} in distribution,
where $\lbrace \lambda_i \rbrace_{i=1}^l$ are the non-zero eigenvalues of $\Omega$, and $s_i$ the corresponding eigenvectors.
\end{theorem}

\begin{proof} To find the limiting distribution of
    \begin{align*}
        \text{CRT}_n = n\cdot  \text{vec}(\hat{U}_2^T \hat{\Pi} \hat{V}_2)^T \text{vec}(\hat{U}_2^T \hat{\Pi} \hat{V}_2),
    \end{align*} we first examine $\text{vec}(\hat{U}_2^T \hat{\Pi} \hat{V}_2)^T$.
    By the delta method,
    \begin{align*}
      \sqrt{n} \cdot \text{vec}(U_2 \hat{\Pi} V_2 - U_2 \Pi V_2) \to \mathcal{N}(0, \Omega).
    \end{align*} Denoting by  $\Sigma_2 \in \mathbb{R}^{k-r \times k-r}$ the submatrix collecting rows and columns $r+1, \dots, k$ of the singular value matrix of $\Theta$, we have $U_2 \Pi V_2 = \sqrt{n}^{-1} \Sigma_2$, which yields
    \begin{align*}
        \sqrt{n} \cdot \text{vec}(U_2 \hat{\Pi} V_2) \to \mathcal{N}(\text{vec}(\Sigma_2), \Omega).
    \end{align*}
    Hence, by \citet[Corollary 1]{Al-Sadoon2017},
    \begin{align*}
     \sqrt{n} \cdot \text{vec}(\hat{U}_2 \hat{\Pi} \hat{V}_2) \to \mathcal{N}(\text{vec}(\Sigma_2), \Omega).
    \end{align*} 
By the spectral theorem, $\Omega$ can be decomposed as $\Omega = SDS^{-1}$ for some orthogonal matrix $S$ and diagonal matrix $D$. Let $X \sim \mathcal{N}(\text{vec}(\Sigma_2), \Omega)$, and let $Y = SX$. Then $Y \sim \mathcal{N}(S\text{vec}(\Sigma_2), D)$ and, in particular, has independent components, so that
\begin{equation*}
  X^T X = X^T S^T S X =  Y^T Y = \sum_{i=1}^k Y_i^2 \sim \sum_{i=1}^k d_i \chi^2\left(\frac{(S\text{vec}(\Sigma_2))_i^2}{d_i}, 1\right). 
\end{equation*}
\end{proof}
\smallskip

It is difficult, however, to map the type of alternative considered in Theorem~\ref{thm:local-power} into an interpretable type of local alternative for our setting. For example, in the alternative (A2) with the additional latent scaled by $\delta$ considered in the main part of the paper, we obtain \begin{equation*} M = \Psi +
   \begin{pmatrix} \delta^2 (b_{02}^{2} \omega_{22}) & \delta^2 (b_{02} b_{12} \omega_{22}) & \delta^2 (b_{12}^{2} \omega_{22}) \\ \delta^3 (b_{02}^{3} \omega_{222}) & \delta^3 (b_{02}^{2} b_{12} \omega_{222}) & \delta^3 (b_{02} b_{12}^{2} \omega_{222}) \\ \delta^3 (b_{02}^{2} b_{12} \omega_{222}) & \delta^3 (b_{02} b_{12}^{2} \omega_{222}) & \delta^3 (b_{12}^{3} \omega_{222}) \end{pmatrix}
\end{equation*}
where $\Psi$ has rank two. Since some of the rows are scaled by $\delta^2$ and some by $\delta^3$, the last singular value of this matrix  admits no simple expression as a function of $\delta$ but is some rational function, which includes powers ranging from $\delta^2$ to $\delta^{16}$.

\subsection{Incomplete U-statistic} \label{subsec:u-stat}

\emph{U-statistics.}  The incomplete U-statistics is a
method to test polynomial (in-)equalities in  parameters of a
multivariate distribution $P^X$. Here, we focus on the case that these
parameters are moments. We write $\mathcal{M}^{\eta}$ for the set of
all moments with order at most $\eta$, and let
$$f = \alpha_0 + \sum_{k=1}^{d} \sum_{\mu_1, \dots, \mu_k \in \mathcal{M}^{\eta}} \alpha_{(\mu_1, \dots, \mu_k)}\mu_1 \cdots \mu_k$$ be an arbitrary polynomial in the moments of $P^X$. For estimating such a polynomial, the plug-in statistic can be biased. For example, the statistic 
$$
S_n = \overline{X_1} \; \overline{X_2} = \frac{n^2-n}{n^2}\sum_{i, j = 1,
i \neq j}^n X_1^{(i)} X_2^{(j)} + \frac{1}{n}\sum_{i=1}^n X_1^{(i)} X_2^{(i)}
$$
for estimating the polynomial $E(X_1^{(1)})E(X_2^{(1)})$ has
expectation $$E\left(S_n\right) = \frac{n-1}{n}
E(X^{(1)}_1)E(X^{(1)}_2) + \frac{1}{n} E\left(X^{(i)}_1
X^{(i)}_2\right).$$ The biasedness originates from the occurrences of
the summands $X^{(i)}_1 X^{(i)}_2$.  However, the polynomials can be
estimated unbiasedly using U-statistics, which build on the
following idea: If a different part of the sample is used to estimate
the factors $\mu_1$ to $\mu_k$ appearing in $f$, then the estimator is
unbiased. We denote by $X^{(l, \dots, m)}$ the subsample of
$\left(X^{(1)}, \dots, X^{(n)}\right)$ consisting of $\left(X^{(l)},
\dots, X^{(m)}\right)$ and by $\hat{\mu}_{i}(X^{(l, \dots m)})$ the
sample moment of $\mu_i$ obtained from the subsample $X^{(l, \dots
m)}$. Define the estimator
\begin{align*}
\breve{h}(X^{(1, \dots, d)}) = \alpha_0 +
\sum_{k=1}^{d} \sum_{\mu_1, \dots, \mu_k \in \mathcal{M}^{\eta}} \alpha_{\mu_1, \dots, \mu_k}\hat{\mu}_1(X^{(1)})\hat{\mu}_2(X^{(2)}) \cdots \hat{\mu}_k(X^{(r)}),
\end{align*} which is unbiased since each $\hat{\mu}_i(X^{(i)})$ is unbiased. Then the U-statistic is obtained  by averaging over all $\breve{h}$ as follows:
\begin{align*}
U_{n}= \frac{1}{\hat{N}} \sum_{i_1, \dots, i_d =1}^n \breve{h}(X^{(i_1, \dots, i_d)}).
\end{align*}
The statistic is often written as an average over the symmetrized versions of $\breve{h}$. Specifically, denoting by
 $I_{n, d} = \{(i_1, \dots, i_d): 1\leq i_1 < \dots < i_d \leq n\}
$  the set containing all ordered subset of size $d$ of $[n]$ and by
\begin{align*}
h(X^{(1, \dots, d)}) = \frac{1}{d!} \sum_{\pi \in S_d} \breve{h}_j(X^{(\pi(1), \dots, \pi(d))}),
\end{align*}
it can be rewritten as $$U_n = \frac{1}{|I_{n,d}|} \sum_{\iota \in I_{n,d}}  h(X^{(\iota)}).$$
While the U-statistic is unbiased, it has two drawbacks: First, it is
computationally expensive since $n^d$ summands occur. Secondly, the
distribution theory supporting the choice of critical values requires
that there do not exist any singularities in the null hypothesis in
the sense that there is no distribution under which  $\text{var}_{x\sim X^{(1)}} \left(E\left(h(x, X^{(2)},
\dots, X^{(p)})\right)\right) = 0$.  The following example shows that
this condition may be violated in our scenario.

\begin{example}
Let $f = \det(A)$ for some matrix $A = (a_{ij})_{i, j \in [d]}\in
\mathbb{R}^{d \times d}$ consisting of moments of $P^X$, and
let $\hat{a}_{i j}(X^{(k)})$ be the corresponding sample moments obtained from the subsample $(X^{(k)})$. Then 
$$\breve{h}(X^{(1, \dots, p)}) = 
\sum_{\pi \in \mathfrak{S}_p} \text{sgn}(\pi) \hat{a}_{i, \pi(i)}(X^{(i)}).$$ So, $X^{(1)}$ is used for estimating the entries in the first row of $M$. Laplace expansion gives $$E\left(\breve{h}(x, X^{(2)}, \dots, X^{(p)})\right) = \sum_{j=1}^p (-1)^{1+j} \hat{a}_{1j}(x)  \det(A_{1j}),$$
where $A_{1j}$ is the matrix obtained from removing the first row and the $j$th column from $A$. If $A$ has rank even lower than $d-1$, then all $\det(A_{1j})$ are zero and so are $E(\breve{h}(x, X^{(2)}, \dots, X^{(p)}))$  and $E(h(x, X^{(2)}, \dots, X^{(p)}))$. 
\end{example}

\emph{Incomplete U-statistic.} These deficiencies can be circumvented
by considering incomplete U-statistics \citep{Sturma}. An incomplete U-statistic is obtained by randomly choosing some of the summands in the U-statistic. Formally, 
\begin{align*}
U_{n,N}= \frac{1}{\hat{N}} \sum_{\iota \in I_{n,d}} Z_{\iota} h(X^{(\iota)}),
\end{align*}
where the computational budget parameter $N$ is lower than or equal to $\binom{n}{d}$, $Z_{\iota}$ are Bernoulli random variables with success probability $N / \binom{n}{d}$ for all $\iota \in I_{n,d}$, and $$\hat{N}=\sum_{\iota \in I_{n,d}}Z_{\iota}$$ is the number of successes. To now test the null hypothesis,
\begin{align*}
H_0: f_i(X) \geq 0 \quad (i = 1, \dots, q),
\end{align*}
consisting of several polynomial constraints, the minimum of the studentized incomplete U-statistics is used as the test statistic. 
So, denoting by $\hat{\sigma}_j$ is an estimate of the variance of $U'_{n, N, j}$, the test statistic is defined as
\begin{align*}
\mathcal{T}_n = \max_{1 \leq j \leq q} \sqrt{n} U'_{n, N, j} / \hat{\sigma}_j.
\end{align*}
The critical values are calibrated by using that under the null hypothesis
\begin{align*}
\mathcal{T}_n \leq \max_{1 \leq j \leq q} \sqrt{n} (U'_{n, N, j} - f_j(\theta)) / \hat{\sigma}_j.
\end{align*}
For the right-hand side, one can show that under mild assumptions, the
distribution of 
\begin{align*}
\max_{1 \leq j \leq q} \sqrt{n} (U'_{n, N, j} - f_j(\theta)) /
\hat{\sigma}_j
\end{align*}
is well approximated by that of $\max_{1 \leq j \leq q} Y_j /
\sigma_j$, where $Y \sim \mathcal{N}_q(0, d^2 W_g + \alpha_n W_h)$, $\alpha_n = n/N$, $W_h = \text{cov}(h(X^{(1, \dots, d)}))$, and $W_g = \text{cov}(g(X^{(1)}))$. The covariance of $Y$ as well as the estimated variances $\hat{\sigma}_j$ are obtained using Bootstrap. For more details, we refer to \cite{Sturma}.

\section{Additional Simulations}
\subsection{Other Choices for the Dimension}
\begin{figure}[t] 
\centering
\begin{subfigure}[c]{0.32\linewidth}
\subcaption{$H_0, p=3$, $l=0$}
\includegraphics[width=\linewidth]{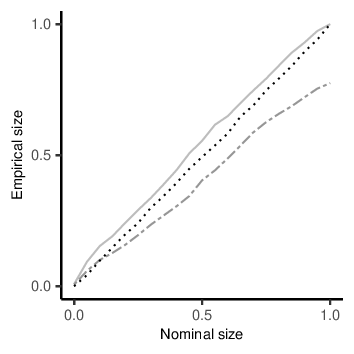}
\end{subfigure}
\begin{subfigure}[c]{0.32\linewidth}
\subcaption{$H_0, p=3$, $l=1$}
\includegraphics[width=\linewidth]{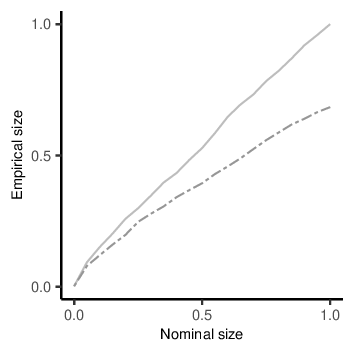}
\end{subfigure}
\begin{subfigure}[c]{0.32\linewidth}
\subcaption{$H_0, p=4$, $l=1$}
\includegraphics[width=\linewidth]{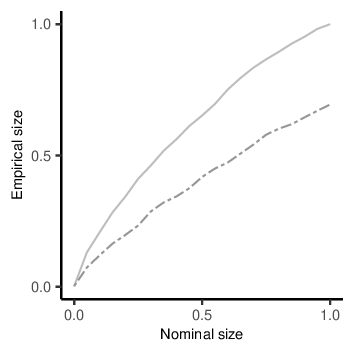}
\end{subfigure}
\begin{subfigure}[c]{0.32\linewidth}
\subcaption{$A1, p=3$, $l=0$}
\includegraphics[width=\linewidth]{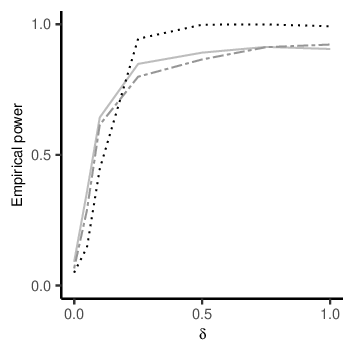}
\end{subfigure}
\begin{subfigure}[c]{0.32\linewidth}
\subcaption{$A1, p=3$, $l=1$}
\includegraphics[width=\linewidth]{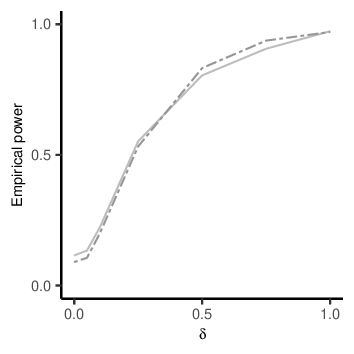}
\end{subfigure}
\begin{subfigure}[c]{0.32\linewidth}
\subcaption{$A1, p=4$, $l=1$}
\includegraphics[width=\linewidth]{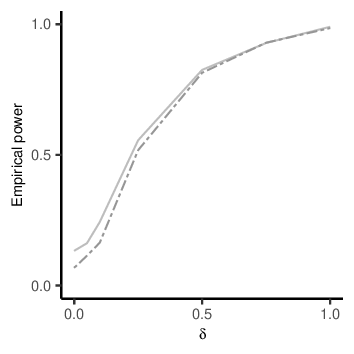}
\end{subfigure}
\begin{subfigure}[c]{0.8\linewidth}
\includegraphics[width=\linewidth]{legend.eps}
\end{subfigure}
\caption{\label{fig:other_choices_p} Also for $p=3, l=0,1$ and $p=4, l=1$, the different tests hold size under $H_0$ and have power against (A1).}
\end{figure}
As displayed in Table 1 in the main paper, our suggested tests draw on different
algebraic information depending on the concerned dimension $p$ and
number of latent variables $l$; confounding in low dimensions requires
us to turn to more refined moment constraints.  The simulations
reported in the main part of the paper treat all those
different cases except for the choices $p=3$ and $l=0$,
 $p=3$ and $l=1$, and $p=4$ and $l=1$. Therefore, we include the simulation results for
those three pairs of $(p,l)$ here; see Figure \ref{fig:other_choices_p}. As in the main paper, we always perform 1000 replications and draw the noise terms from a Gamma distribution.

 \subsection{Higher Number of Latents}
\begin{figure}[t] 
  \centering
  \begin{subfigure}[c]{0.32\linewidth}
  \subcaption{$H_0$, $p=2, l=2$}
  \includegraphics[width=\linewidth]{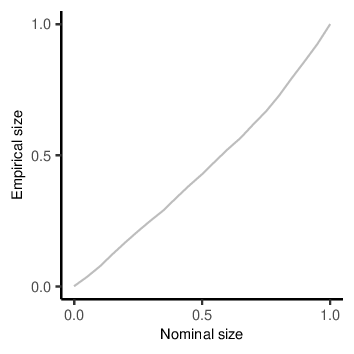}
  \end{subfigure}
  \begin{subfigure}[c]{0.32\linewidth}
  \subcaption{A1, $p=2, l=2$}
  \includegraphics[width=\linewidth]{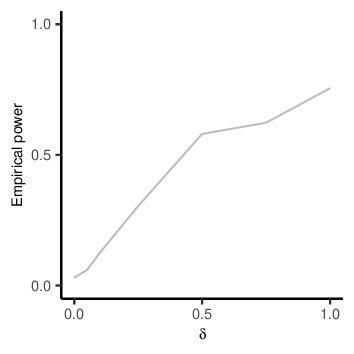}
  \end{subfigure}
  \begin{subfigure}[c]{0.32\linewidth}
  \subcaption{A2, $p=2, l=2$}
  \includegraphics[width=\linewidth]{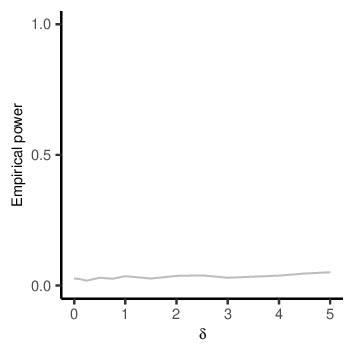}
  \end{subfigure}
  \begin{subfigure}[c]{0.32\linewidth}
  \subcaption{$H_0$, $p=4, l=2, 3, 4$}
  \includegraphics[width=\linewidth]{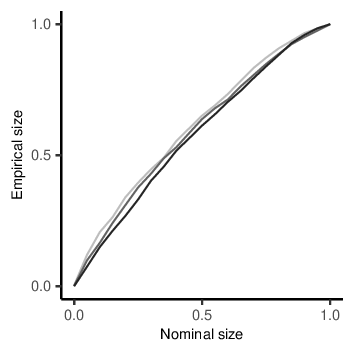}
  \end{subfigure}
  \begin{subfigure}[c]{0.32\linewidth}
  \subcaption{A1, $p=4, l=2, 3, 4$}
  \includegraphics[width=\linewidth]{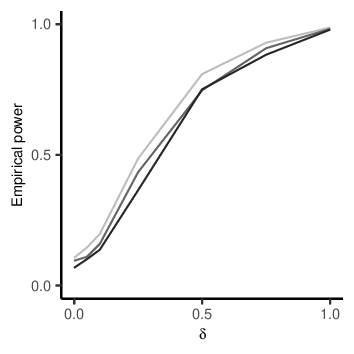}
  \end{subfigure}
  \begin{subfigure}[c]{0.32\linewidth}
  \subcaption{A2, $p=4, l=2, 3, 4$}
  \includegraphics[width=\linewidth]{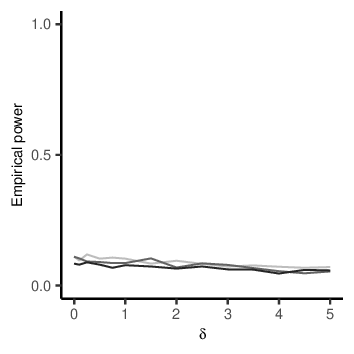}
  \end{subfigure}
  \begin{subfigure}[c]{0.3\linewidth}
  \centering
  \includegraphics[height=.17cm]{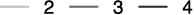}
  \end{subfigure}\\
  \begin{subfigure}[c]{0.32\linewidth}
  \subcaption{$H_0$, $p=6, l=2, \dots, 6$}
  \includegraphics[width=\linewidth]{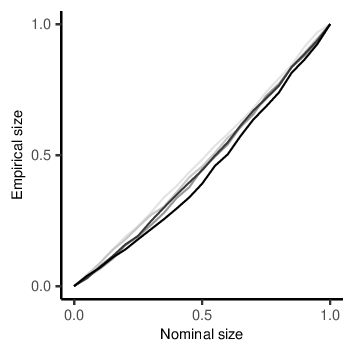}
  \end{subfigure}
  \begin{subfigure}[c]{0.32\linewidth}
  \subcaption{A1, $p=6, l=2, \dots, 6$}
  \includegraphics[width=\linewidth]{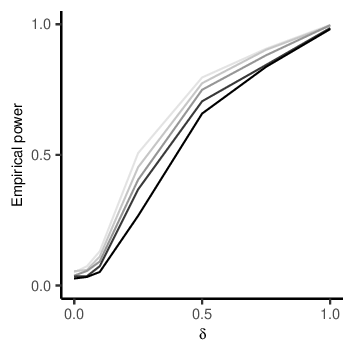}
  \end{subfigure}
  \begin{subfigure}[c]{0.32\linewidth}
  \subcaption{A2, $p=6, l=2, \dots, 6$}
  \includegraphics[width=\linewidth]{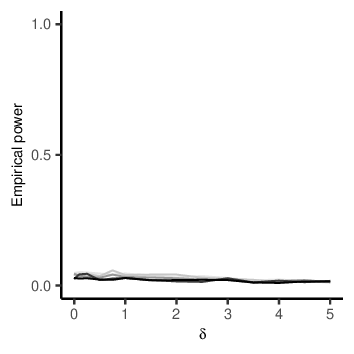}
  \end{subfigure}
  \begin{subfigure}[c]{0.4\linewidth}\centering
  \includegraphics[height=.33cm]{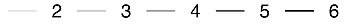}
  \end{subfigure}
  \caption{\label{fig:higher_l} The test shows only a slight decrease in performance when confronted with higher $l$, as exemplified here for $p=2,4$ with $n=1000$, and $p=6$ using $n=2000$.}
  \end{figure}

 To examine whether the tests are still effective when confronted with a higher number of latents $l$, we conduct additional simulations with $p=2, 4, 6$ and $l$ ranging from $2$ to $p$. Figure \ref{fig:higher_l} illustrates that the performance exhibits only a marginal decrease as the number of latents increases.
  
\subsection{Gaussian Noise Terms}

Next, we consider Gaussian noise terms. In this case, the matrix
$M^{(k_1, \dots, k_2)}$ has only rank one in the population. For the incomplete U-statistic,  rank one corresponds to a degenerate null hypothesis in the sense that $\text{var}(g)=0$. As discussed in Section~\ref{sec:statistical_methods}, in this degenerate case, there are still theoretical guarantees that the test holds the level for the incomplete U-statistic but not for the CR statistic. 
\begin{figure}[t] 
\centering
\begin{subfigure}[c]{0.32\linewidth}
\subcaption{$H_0, p=2$}
\includegraphics[width=\linewidth]{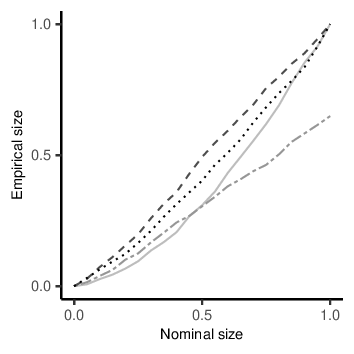}
\end{subfigure}
\begin{subfigure}[c]{0.32\linewidth}
\subcaption{$H_0,p=4$}
\includegraphics[width=\linewidth]{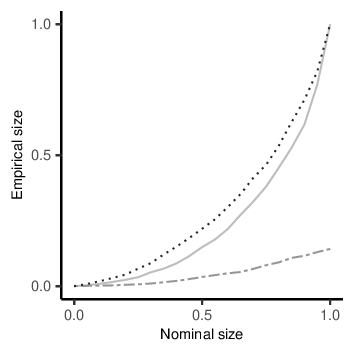}
\end{subfigure}
\begin{subfigure}[c]{0.32\linewidth}
\subcaption{$H_0,p=20$}
\includegraphics[width=\linewidth]{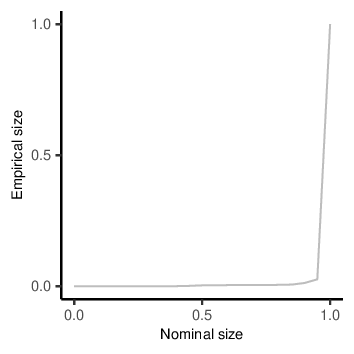}
\end{subfigure}
\begin{subfigure}[c]{0.32\linewidth}
\subcaption{$A1, p=2$}
\includegraphics[width=\linewidth]{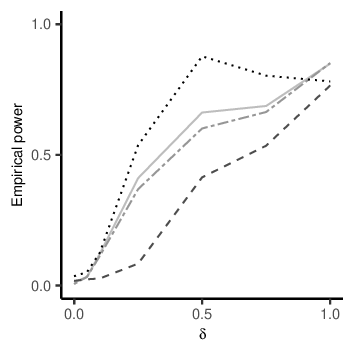}
\end{subfigure}
\begin{subfigure}[c]{0.32\linewidth}
\subcaption{$A1,p=4$}
\includegraphics[width=\linewidth]{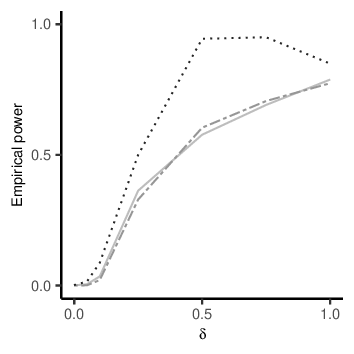}
\end{subfigure}
\begin{subfigure}[c]{0.32\linewidth}
\subcaption{$A1,p=20$}
\includegraphics[width=\linewidth]{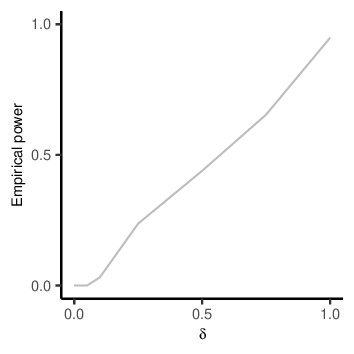}
\end{subfigure}
\begin{subfigure}[c]{0.8\linewidth}
\includegraphics[width=\linewidth]{legend.eps}
\end{subfigure}
\caption{\label{fig:H_O_gaussian} Empirical sizes under $H_0$, and empricial powers against $(A1)$  with Gaussian noise.}
\end{figure}
Figure \ref{fig:H_O_gaussian} displays our simulation results for data simulated according to the null hypothesis as described in the main paper, except that the noise terms are now drawn from a Gaussian distribution with mean $0$ and standard deviation uniformly chosen from $[1/2, 2]$. Unsurprisingly, the incomplete U-statistic outperforms
the other methods in terms of holding the size, while the test based on the CR
statistic is too conservative.  However, in terms of power
to detect non-linear relations between the variables,
Gaussian noise terms do not pose a problem as the joint distribution
$P^X$ is then still non-Gaussian.
Hence, $M^{(k_1, \dots, k_2)}$ has full rank such that
the CR statistic is able to detect the non-linearity. The subsequent
experimental setup shown in the second row of Figure \ref{fig:H_O_gaussian} indeed demonstrates that also in the case of
Gaussian noise,
the test exhibits power against alternatives; even though somewhat higher
sample sizes seem to be needed to reach  power comparable to that seen
for cases with
non-Gaussian noise.

\subsection{Further Alternatives} 
\begin{figure}[htbp]
    \centering
    \begin{subfigure}[b]{0.4\textwidth}
        \subcaption{\(p=2, l = 0, n = 1000\)}
        \includegraphics[width=\textwidth]{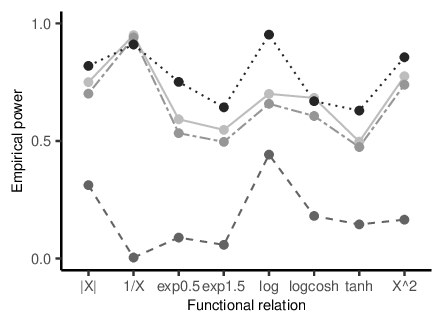}
    \end{subfigure}
    \begin{subfigure}[b]{0.4\textwidth}
        \subcaption{\(p=2, l = 1, n = 1000\)}
        \includegraphics[width=\textwidth]{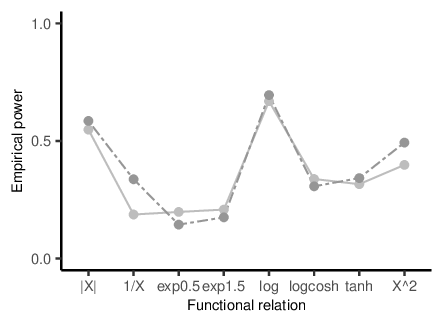}
    \end{subfigure}
    \\
    \begin{subfigure}[b]{0.4\textwidth}
        \subcaption{\(p=2, l = 1, n = 2000\)}
        \includegraphics[width=\textwidth]{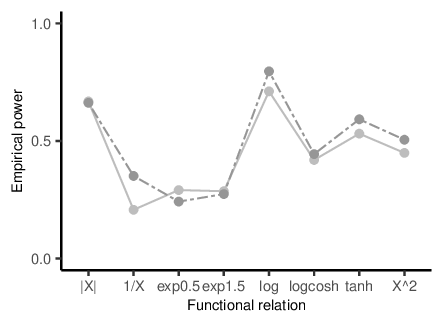}
    \end{subfigure}
    \begin{subfigure}[b]{0.4\textwidth}
        \subcaption{\(p=2, l = 1, n = 4000\)}
        \includegraphics[width=\textwidth]{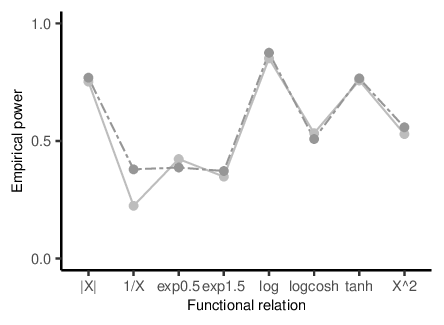}
    \end{subfigure}
\begin{subfigure}[c]{0.8\linewidth}
\includegraphics[width=\linewidth]{legend.eps}
\end{subfigure}
    \caption{Empirical power against further non-linear alternatives. }
    \label{fig:further_alternatives}
\end{figure}

To shed light on the performance against further types of alternatives, 
following \cite{Biza_2020}, we consider different instances of the post-nonlinear model \citep{Zhang_2009}. Specifically, we sample variables $X_1, \dots, X_{p+l}$ as shown in Table \ref{table:furter_alternatives}. Then, if $l>0$, the first $l$ columns are dropped from the resulting data matrix. In all cases, the remaining  columns are randomly permuted to ensure a random causal order. Note that this approach samples based on directed acyclic graphs rather than complete graphs, as done previously. Figure \ref{fig:further_alternatives} shows that, similar to the other alternatives, the incomplete $U$-statistic used alone achieves the lowest power. For $l=0$, there are notable differences between the various functional relations, with exp0.5, exp1.5, and tanh appearing to be the most challenging ones. This is likely linked to the different data-generating mechanisms yielding substantially different higher-order moments. For $l=1$, higher sample sizes are needed to reach the same power level seen for $l=0$, which is unsurprising given that the test depends on higher-order cumulants, specifically orders 3, 4, and 5, instead of 2 and 3.
\renewcommand{\arraystretch}{1.5}
\begin{table}[t]
\tbl{Further alternatives considered: We sample data according to the post-nonlinear model, via the following relations.}
{\begin{tabular}{clcl}
 & Functional relations &  &  Functional relations \\
tanh & $X_i = \tanh \left(\sum_{j=1}^{i-1}\left(b_{ij} X_j\right) + \epsilon_i \right)$ & $X^2$ & $X_i = \sum_{j=1}^{i-1}\left(b_{ij} X_j^2 \right) + \epsilon_i$ \\
$1 / X$ & $X_i = \sum_{j=1}^{i-1}\left(b_{ij} X_j^{-1} \right) + \epsilon_i$ & $|X|$ & $X_j = \sum_{j=1}^{i-1}\left(b_{ij} |X_j|\right) + \epsilon_i$ \\
exp0.5 & $X_i = \sum_{j=1}^{i-1}\left(b_{ij} \operatorname{sign}(X_j) |X_j|^{0.5}\right) + \epsilon_i$ & exp1.5 & $X_i = \sum_{j=1}^{i-1}\left(b_{ij} \operatorname{sign}(X_j) |X_j|^{1.5} \right) + \epsilon_i$ \\
$\log$ & $X_i = \sum_{j=1}^{i-1}\left(b_{ij} \log \left| X_j \right|\right) + \epsilon_i$ & logcosh & $X_i = \sum_{j=1}^{i-1}\left(b_{ij} \log \left( \cosh(X_j) \right) \right) + \epsilon_i$ \\
\end{tabular}}
\label{table:furter_alternatives}
\end{table}
\renewcommand{\arraystretch}{1}

\subsection{Choice of Orders}
\begin{figure}[t] 
  \centering
  \begin{subfigure}[c]{0.32\linewidth}
  \subcaption{$H_0$, varying $k_2$}
  \includegraphics[width=\linewidth]{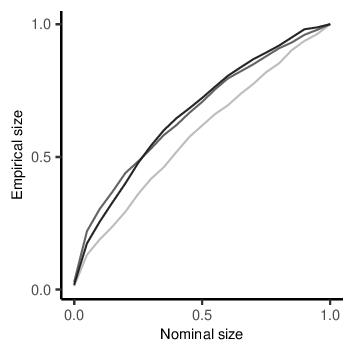}
  \end{subfigure}
  \begin{subfigure}[c]{0.32\linewidth}
  \subcaption{A1, varying $k_2$}
  \includegraphics[width=\linewidth]{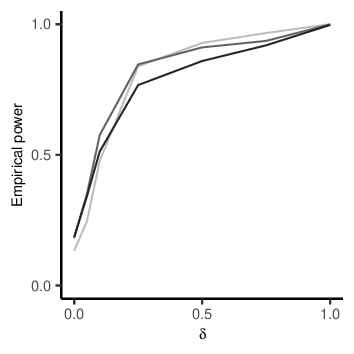}
  \end{subfigure}
  \begin{subfigure}[c]{0.32\linewidth}
  \subcaption{A2, varying $k_2$}
  \includegraphics[width=\linewidth]{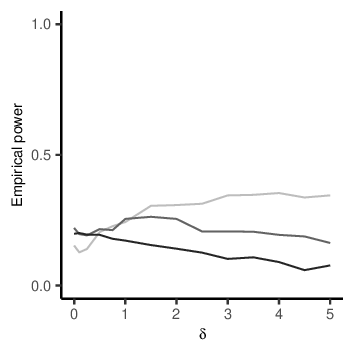}
  \end{subfigure}
  \begin{subfigure}[c]{0.28\linewidth}
  \includegraphics[width=\linewidth]{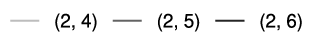}
  \end{subfigure}\\
  \begin{subfigure}[c]{0.32\linewidth}
  \subcaption{$H_0$, varying $(k_1, k_2)$}
  \includegraphics[width=\linewidth]{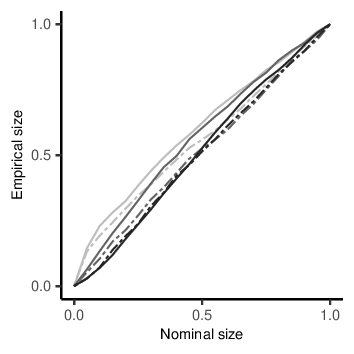}
  \end{subfigure}
  \begin{subfigure}[c]{0.32\linewidth}
  \subcaption{A1, varying $(k_1, k_2)$}
  \includegraphics[width=\linewidth]{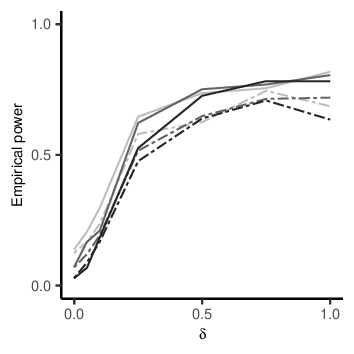}
  \end{subfigure}
  \begin{subfigure}[c]{0.32\linewidth}
  \subcaption{A2, varying $(k_1, k_2)$}
  \includegraphics[width=\linewidth]{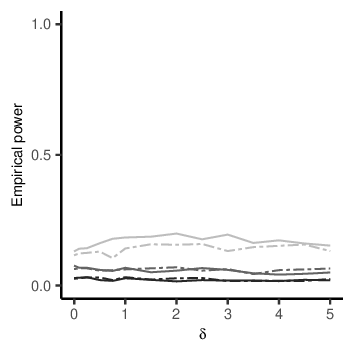}
  \end{subfigure}
  \begin{subfigure}[c]{0.6\linewidth}
  \includegraphics[width=\linewidth]{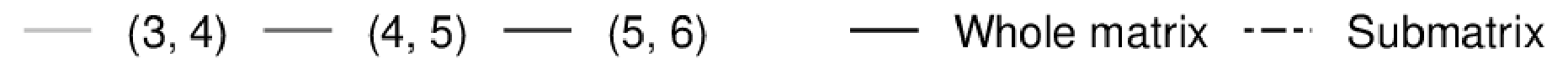}
  \end{subfigure}
  \caption{\label{fig:varying_k_2} Lower choices of orders $k_1, k_2$ in $M^{(k_1, \dots, k_2)}$ tend to yield a better performance. }
  \end{figure}

Lastly, we investigate how the choice of orders $k_1, k_2$ in Theorem 1 affects the tests' performance. We focus on the case $p=2$ and $l=0$ as increasing the orders significantly slows down the computation time in a simulation study. First, we fix $k_1 = 2$ and vary $k_2$ across $k_2 = 4,5,6$. This variation affects not only the orders but also increases the number of rows and columns of $M^{(k_1, \dots, k_2)}$, which could potentially impact the performance of the test of rank. Therefore, in a second experiment, we vary both parameters $k_1, k_2$ jointly, while fixing their difference. Specifically, we test the pairs $(k_1, k_2) = (3, 4), (4, 5), (5, 6)$, and run the simulations twice for each choice: once, we use the full matrix $M^{(k_1, \dots, k_2)}$ and once with the upper right submatrix of $M^{(k_1, \dots, k_2)} \in \mathbb{R}^{3\times 3}$, which has the same size as $M^{(2,3)}$. Overall, the lower orders tend to yield a better performance, while selecting the submatrix slightly decreases the performance, as shown in Figure \ref{fig:varying_k_2}.

\section{Results of all Methods on the T\"ubingen Cause-Effect-Pairs}
\begin{figure}[t]
\centering
\includegraphics[width=.98\textwidth]{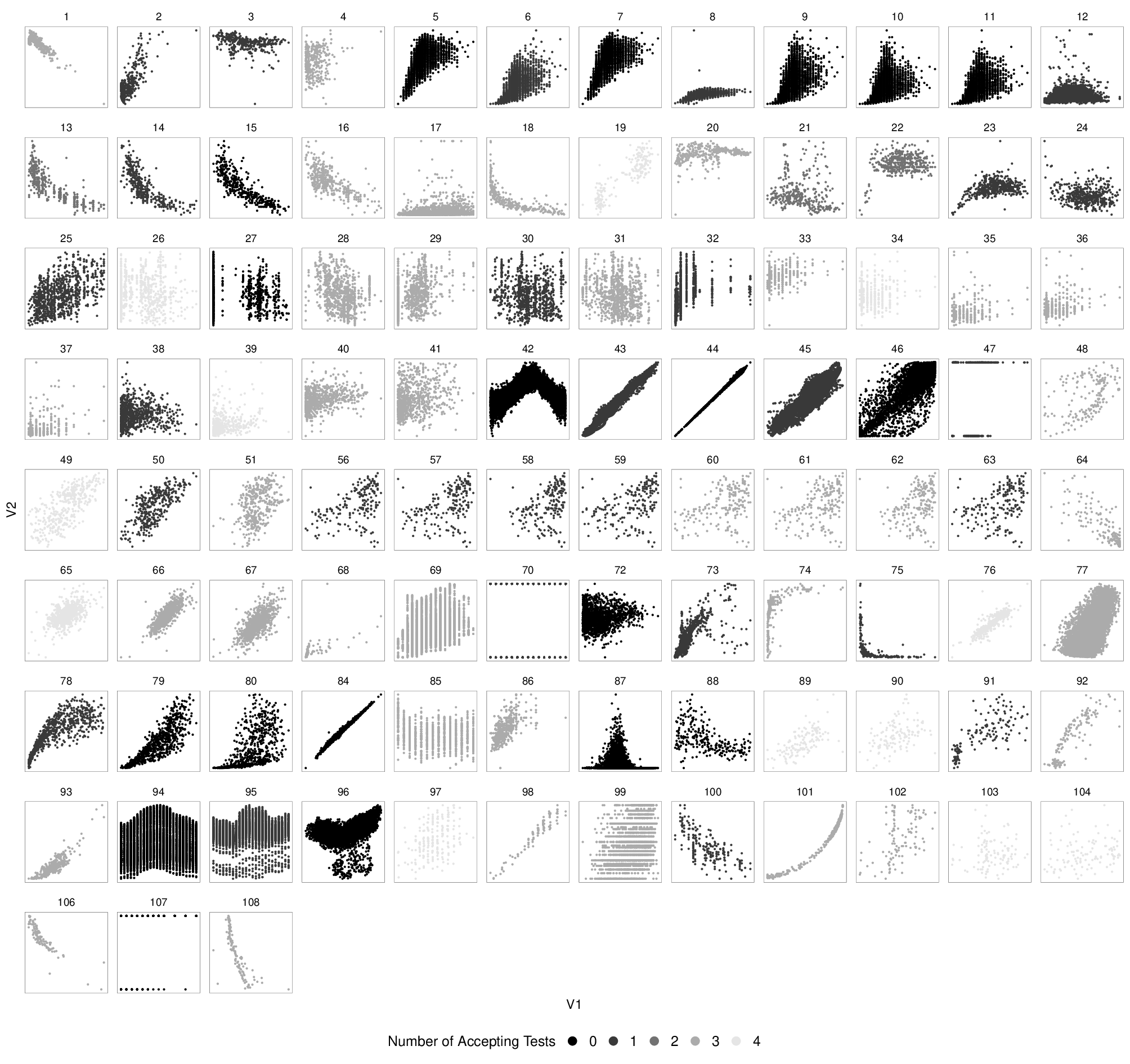}
\caption{Number of test that accepted the model without latents for each pair.}\label{fig:tueb_all_l=0}
\end{figure}
We apply all test for $p=2$ and $l=0$ to the T\"ubingen
cause-effect-pairs. The test dCovICA accepts 13 pairs, the CR statistic 50, the combination 53 and the U-statistic used alone 77. Figure \ref{fig:tueb_all_l=0} indicates the number of tests accepting each pair. In most cases, if a test accepts a pair, then the less conservative tests also do. The only exceptions are:
Pair 70 is accepted by dCovICA, not by any other test, Pair 21 is accepted solely by the CR statistic and the combined test, and Pair 57 is accepted only by the combined test.

\begin{figure}[t]
\centering
\includegraphics[width=.98\textwidth]{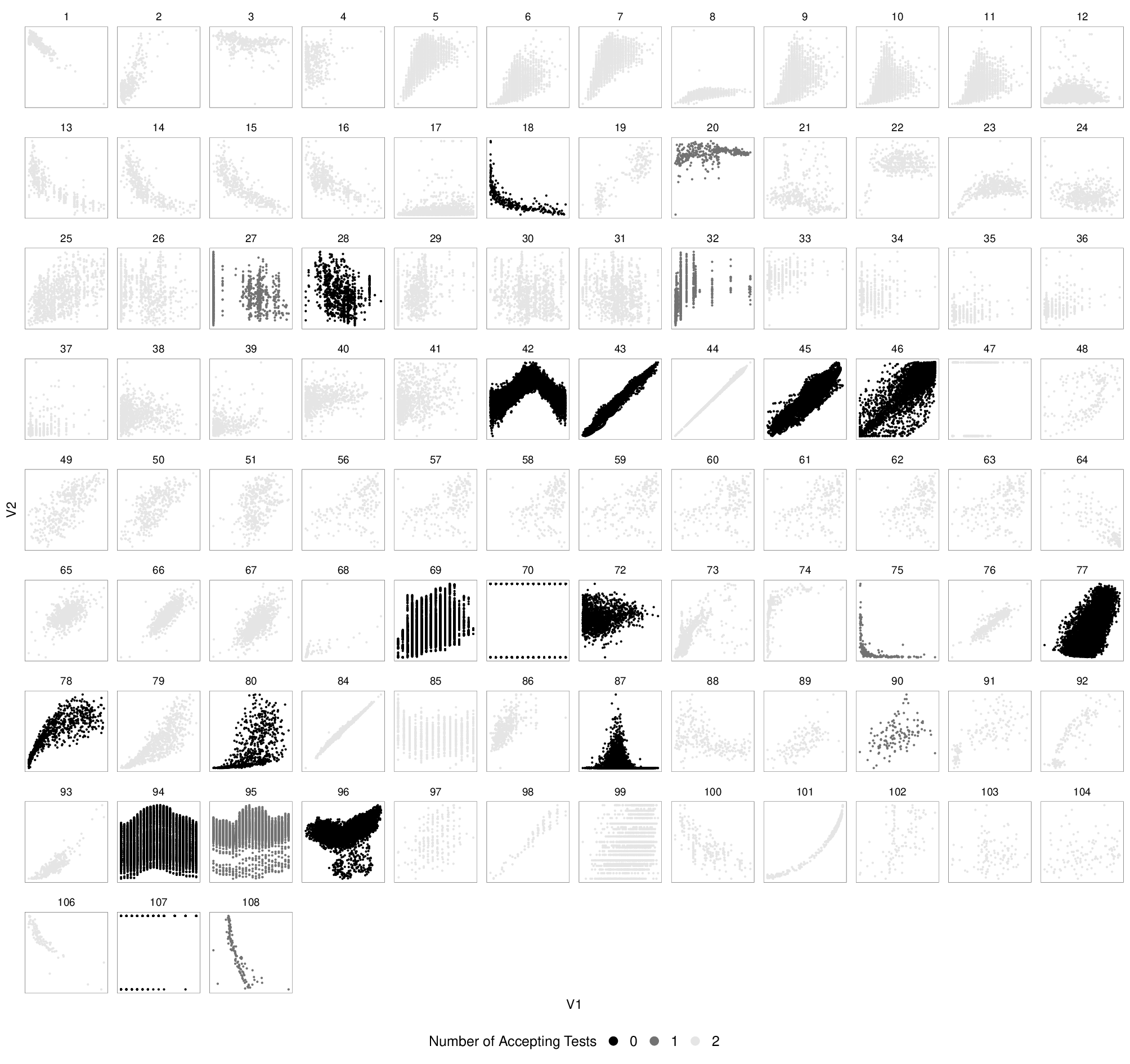}
\caption{Number of test that accepted the model with one latent for each pair.}\label{fig:tueb_all_l=1}
\end{figure}

Figure \ref{fig:tueb_all_l=1} reports the number of tests accepting the null hypothesis with $l=1$. The CR statistic applied with both rank conditions is slightly more conservative than the CR statistic taking only the condition on $M^{(3,4,5)}$ into account, with the former accepting 78, the latter 81 pairs. Five pairs are accepted by the CR statistic with just  $M^{(3,4,5)}$, but not by the combined test, namely 20, 32, 90, 95, 108.

%\bibliography{paper-ref}

\end{document}